\documentclass[a4paper]{jpconf-mod}
\usepackage{graphicx,iopams}
\usepackage{amsmath,amssymb}
\usepackage[hyperindex,breaklinks]{hyperref}
\usepackage{url,cite} 

\DeclareGraphicsExtensions{.pdf,.png}
\begin{document}
\markright{IFIC/17-14} %IFIC/17-14
\title[The MoEDAL experiment at the LHC: status and results]{The MoEDAL experiment at the LHC: \\ status and results}

\author{Vasiliki A Mitsou \\ on behalf of the MoEDAL Collaboration}

\address{Instituto de F\'isica Corpuscular (IFIC), CSIC -- Universitat de Val\`encia, \\
C/ Catedr\'atico Jos\'e Beltr\'an 2, E-46980 Paterna (Valencia), Spain}

\ead{vasiliki.mitsou@ific.uv.es}

\begin{abstract}
The MoEDAL experiment at the LHC is optimised to detect highly ionising particles such as magnetic monopoles, dyons and (multiply) electrically charged stable massive particles predicted in a number of theoretical scenarios. MoEDAL, deployed in the LHCb cavern, combines passive nuclear track detectors with magnetic monopole trapping volumes (MMTs), while spallation-product backgrounds are being monitored with an array of MediPix pixel detectors. An introduction to the detector concept and its physics reach, complementary to that of the large general purpose LHC experiments ATLAS and CMS, will be given. Emphasis is given to the recent MoEDAL results at 13~TeV, where the null results from a search for magnetic monopoles in MMTs exposed in 2015 LHC collisions set the world-best limits on particles with magnetic charges more than 1.5~Dirac charge. The potential to search for heavy, long-lived supersymmetric electrically-charged particles is also discussed.
\end{abstract}

%%%%%%%%%%%%%%%%%%%%%%%%%%%%%%%%%%%%%%%%%%%%%%%%%%%
%%%%%%%%%%%%%%%%%%%%%%%%%%%%%%%%%%%%%%%%%%%%%%%%%%%
\section{Introduction}\label{sc:intro}

MoEDAL (Monopole and Exotics Detector at the LHC)~\cite{moedal-web,moedal-tdr}, the $7^{\rm th}$ experiment at the Large Hadron Collider (LHC)~\cite{LHC}, was approved by the CERN Research Board in 2010. It is designed to search for manifestations of new physics through highly-ionising particles in a manner complementary to ATLAS and CMS~\cite{DeRoeck:2011aa}. The most important motivation for the MoEDAL experiment is to pursue the quest for magnetic monopoles and dyons at LHC energies. Nonetheless the experiment is also designed to search for any massive, stable or long-lived, slow-moving particles~\cite{Fairbairn07} with single or multiple electric charges arising in many scenarios of physics beyond the Standard Model (SM). A selection of the physics goals and their relevance to the MoEDAL experiment are described here and elsewhere~\cite{creta2016}. For an extended and detailed account of the MoEDAL discovery potential, the reader is referred to the \emph{MoEDAL Physics Review}~\cite{Acharya:2014nyr}. Emphasis is given here on recent MoEDAL results, based on the exposure of magnetic monopole trapping volumes to 7-TeV and 8-TeV proton-proton collisions.

The structure of this paper is as follows. Section~\ref{sc:detector} provides a brief description of the MoEDAL detector. Magnetic monopoles and monopolia are briefly discussed in Section~\ref{sc:mm}, whilst Section~\ref{sc:lightsearch} presents the MoEDAL results on monopole searches. Section~\ref{sc:susy} is dedicated to supersymmetric models predicting massive (meta)stable states. Scenarios with doubly-charged Higgs bosons and their observability in MoEDAL are highlighted in Section~\ref{sc:lrsm}. Highly-ionising exotic structures in models with extra spatial dimensions, namely microscopic black holes and D-matter, relevant to MoEDAL are briefly mentioned in Sections~\ref{sc:bh} and Sections~\ref{sc:dmatter}, respectively. The paper concludes with a summary and an outlook in Section~\ref{sc:summary}.

%%%%%%%%%%%%%%%%%%%%%%%%%%%%%%%%%%%%%%%%%%%%%%%%%%%
%%%%%%%%%%%%%%%%%%%%%%%%%%%%%%%%%%%%%%%%%%%%%%%%%%%
\section{The MoEDAL detector}\label{sc:detector}

\begin{figure}[htb]
\begin{center}
\includegraphics[width=0.7\textwidth]{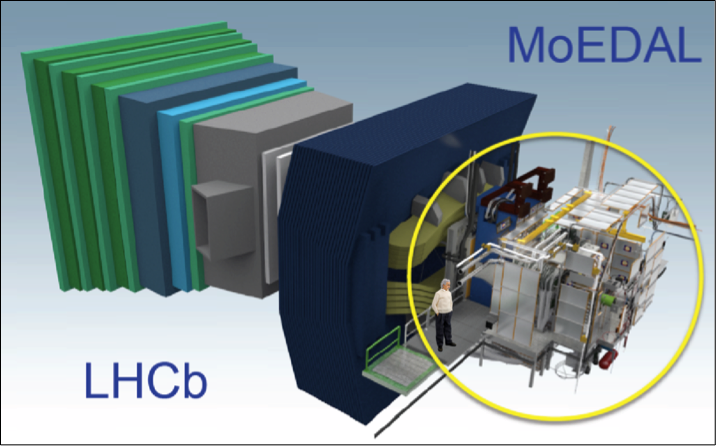}
\caption{ A three-dimensional schematic view of the MoEDAL detector (on the right) around the LHCb VELO region at Point~8 of the LHC.}
\label{Fig:moedal-lhcb}
\end{center}
\end{figure}

The MoEDAL detector~\cite{moedal-tdr} is deployed around the intersection region at Point~8 of the LHC in the LHCb experiment Vertex Locator (VELO)~\cite{LHCb-detector} cavern. A three-dimensional depiction of the MoEDAL experiment is presented in Fig.~\ref{Fig:moedal-lhcb}. It is a unique and largely passive LHC detector comprised of four sub-detector systems. 

%%%%%%%%%%%%%%%%%%%%%%%%%%%%%%%%%%%%%%%%%%%%%%%%%%%
\subsection{Low-threshold nuclear track detectors}\label{sc:ndt}

\begin{figure}[htb]
\begin{minipage}[b]{0.59\textwidth}
\includegraphics[width=\textwidth]{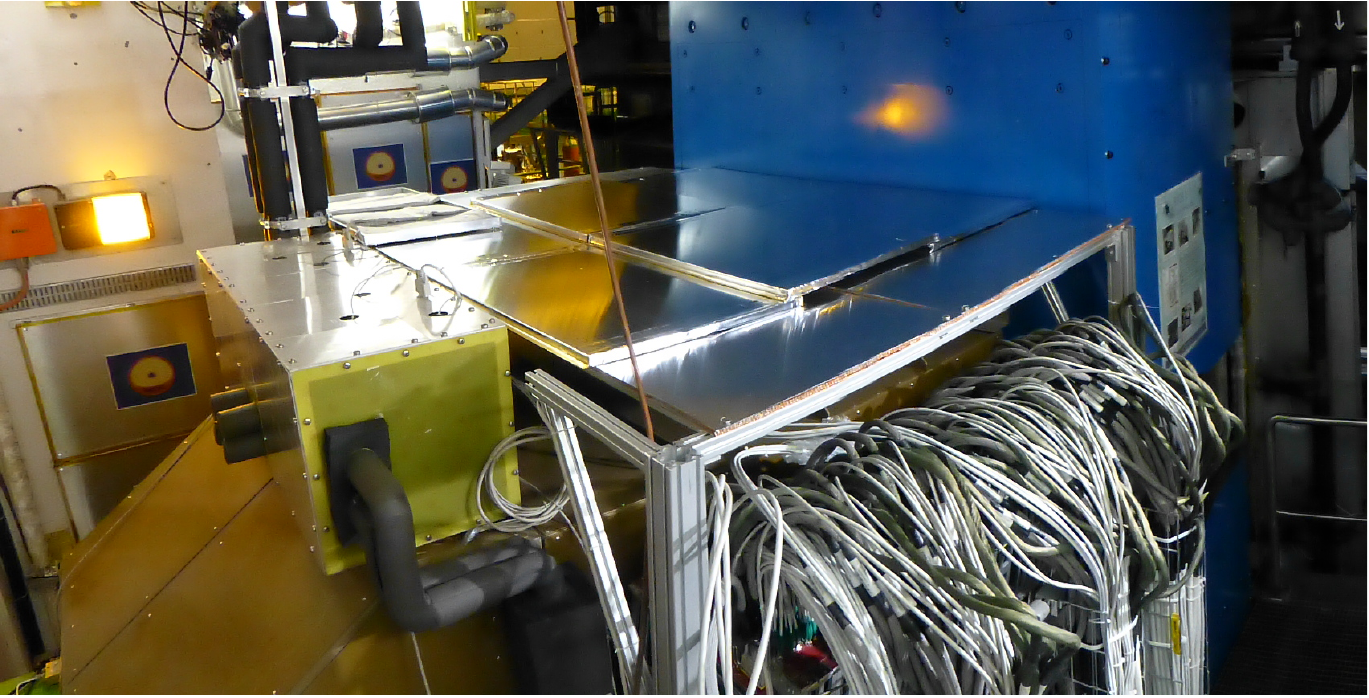}
\caption{\label{fg:ntd}Part of the Run~2 NTD deployment on top of the LHCb VELO.}
\end{minipage}\hspace{2pc}%
\begin{minipage}[b]{0.35\textwidth}
\includegraphics[width=\textwidth]{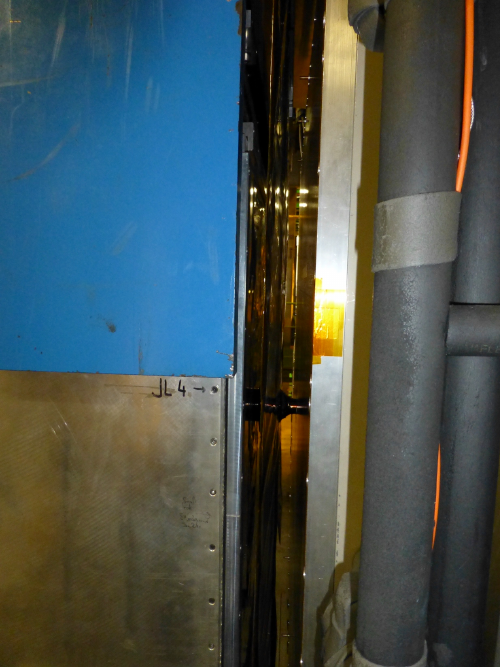}
\caption{\label{fg:vhcc}The VHCC between RICH1 and TT installed for Run~2.}
\end{minipage} 
\end{figure}

The main sub-detector system is made of a large array of CR39\textregistered,  Makrofol\textregistered\ and Lexan\textregistered\ nuclear track detector (NTD) stacks surrounding the intersection area. The passage of a highly-ionising particle through the plastic detector is marked by an invisible damage zone along the trajectory. The damage zone is revealed as a cone-shaped etch-pit when the plastic detector is etched using a hot sodium hydroxide solution. Then the sheets of plastics are scanned looking for aligned etch pits in multiple sheets. The MoEDAL NTDs have a threshold of $Z/\beta\sim5$, where $Z$ is the charge and $\beta=v/c$ the velocity of the incident particle. In proton-proton collision running, the only source of known particles that are highly ionising enough to leave a track in MoEDAL NTDs are spallation products with range that is typically much less than the thickness of one sheet of the NTD stack. In that case the ionising signature will be that of a very low-energy electrically-charged \emph{stopped} particle. This signature is distinct to that of a \emph{penetrating} electrically or magnetically charged particle that will usually traverse every sheet in a MoEDAL NTD stack, accurately demarcating a track that points back to the collision point with a resolution of $\sim 1~{\rm cm}$. The part of the Run-2 NTD deployment which rests on top of the LHCb VELO is visible in Fig.~\ref{fg:ntd}. This is the closest possible location to the interaction point and represents a novelty of this run with respect to earlier installations during Run-1. 

%%%%%%%%%%%%%%%%%%%%%%%%%%%%%%%%%%%%%%%%%%%%%%%%%%%
\subsection{Very high-charge catcher}\label{sc:vhcc}

Another new feature of the Run-2 deployment is the installation of a high-threshold NTD array ($Z/\beta\sim50$): the Very High Charge Catcher (VHCC).  The VHCC sub-detector, consisting of two flexible low-mass stacks of Makrofol\textregistered\ in an aluminium foil envelope, is deployed in the forward acceptance of the LHCb experiment between the LHCb RICH1 detector and the Trigger Tracker (TT), as shown in Fig.~\ref{fg:vhcc}. It is the only NTD (partly) covering the forward region, adding only $\sim0.5\%$ to the LHCb material budget while enhancing considerably the overall geometrical coverage of MoEDAL NTDs.

%%%%%%%%%%%%%%%%%%%%%%%%%%%%%%%%%%%%%%%%%%%%%%%%%%%
\subsection{Magnetic trappers}\label{sc:mmt}

\begin{figure}[htb]
\begin{minipage}[b]{0.46\textwidth}
\includegraphics[width=\textwidth]{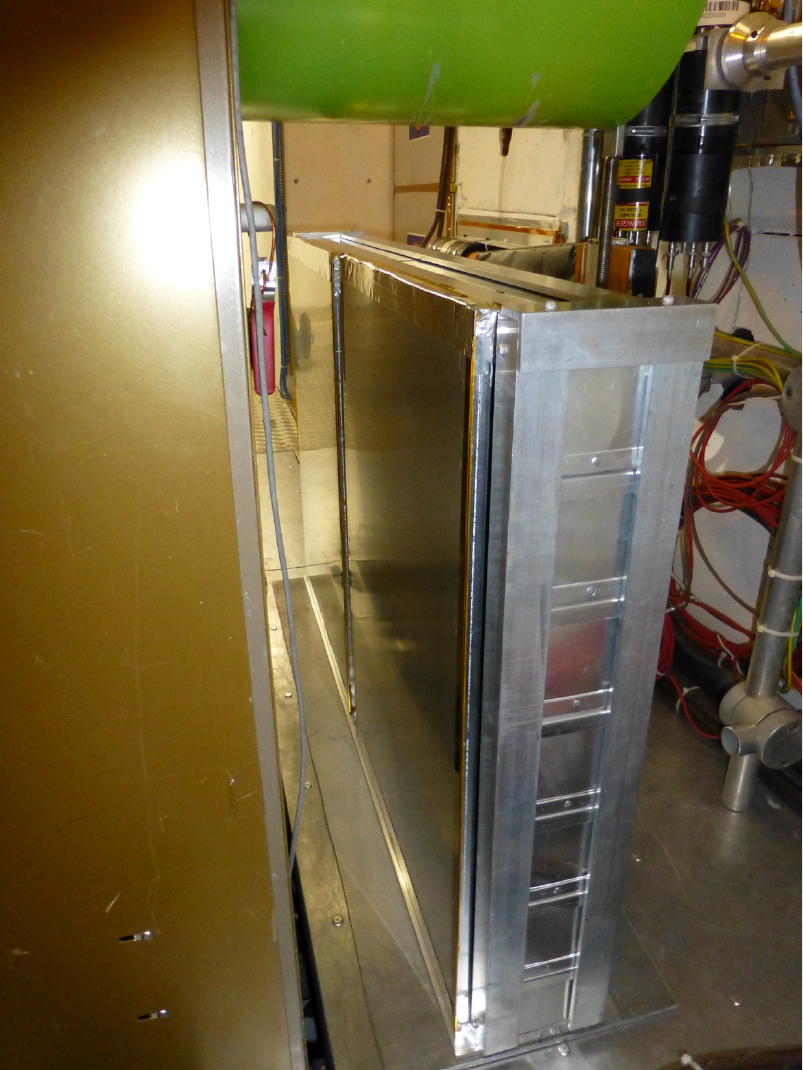}
\caption{\label{fg:mmt}Deployment of the MMT for the LHC Run~2.}
\end{minipage}\hspace{2pc}%
\begin{minipage}[b]{0.46\textwidth}
\includegraphics[width=\textwidth]{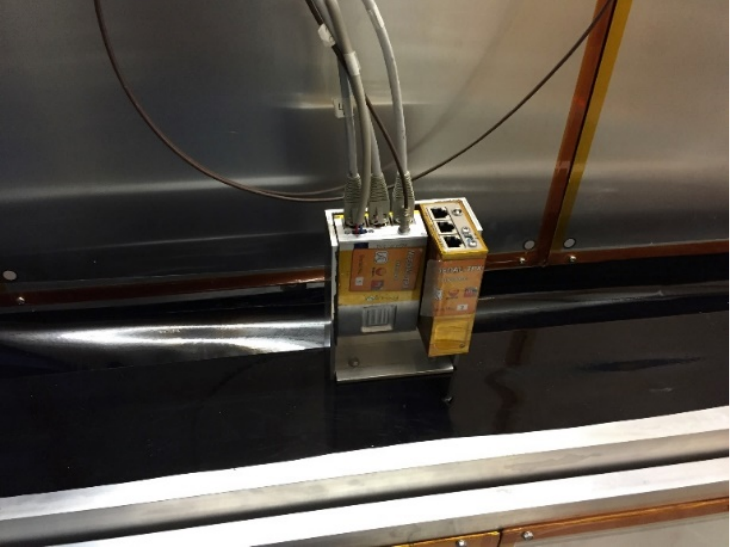}
\caption{\label{fg:timpix}Run~2 deployment of TimePix chips in MoEDAL.}
\end{minipage} 
\end{figure}

A unique feature of the MoEDAL detector is the use of paramagnetic magnetic monopole trappers (MMTs) to capture electrically- and magnetically-charged highly-ionising particles. Such volumes installed in IP8 for the 2015 proton-proton collisions is shown in Fig.~\ref{fg:mmt}. The aluminium absorbers of MMTs are subject to an analysis looking for magnetically-charged particles at a remote SQUID magnetometer facility~\cite{Joergensen:2012gy,DeRoeck:2012wua}. The search for the decays of long-lived electrically charged particles that are stopped in the trapping detectors will subsequently be carried out at a remote underground facility. 

A trapping detector prototype was exposed to 8~TeV proton-proton collisions for an integrated luminosity of 0.75~fb$^{-1}$ in 2012. It comprised an aluminium volume consisting of 11~boxes each containing 18~cylindrical rods of 60~cm length and 2.5~cm diameter. For the 2015 run at 13~TeV, the MMT was upgraded to an array consisting of 672 square aluminium rods with dimension $19\times2.5\times2.5~{\rm cm}^3$ for a total mass of 222~kg in 14~stacked boxes that were placed 1.62~m from the IP8 LHC interaction point under the beam pipe on the side opposite to the LHCb detector. The results for both aforementioned configurations and energies, interpreted in terms of monopole mass and magnetic charge, are presented in Section~\ref{sc:lightsearch}.

%%%%%%%%%%%%%%%%%%%%%%%%%%%%%%%%%%%%%%%%%%%%%%%%%%%
\subsection{TimePix radiation monitors}\label{sc:timepix}

The only non-passive MoEDAL sub-detector system comprises an array of TimePix pixel device arrays ($256\times256$ square pixels with a pitch of $55~{\rm \mu m}$)  distributed  throughout  the  MoEDAL cavern at IP8, forming a real-time radiation monitoring system of highly-ionising beam-related  backgrounds. A photo of its readout setup for the 2015 installations is shown in Fig.~\ref{fg:timpix}. Each pixel of the innovative TimePix chip comprises a preamplifier, a discriminator with threshold adjustment, synchronisation logic and a 14-bit counter. The operation of TimePix in time-over-threshold mode allows a 3D mapping of the charge spreading effect in the whole volume of the silicon sensor, thus differentiating between different types of particles species from mixed radiation fields and measuring their energy deposition~\cite{timepix}.

%%%%%%%%%%%%%%%%%%%%%%%%%%%%%%%%%%%%%%%%%%%%%%%%%%%
%%%%%%%%%%%%%%%%%%%%%%%%%%%%%%%%%%%%%%%%%%%%%%%%%%%
\section{Magnetic monopoles}\label{sc:mm}

The MoEDAL detector is designed to fully exploit the energy-loss mechanisms of magnetically charged particles~\cite{Dirac1931kp,Diracs_idea,tHooft-Polyakov,mm}  in order to optimise its potential to discover these messengers of new physics. There are various theoretical scenarios in which magnetic charge would be produced  at the LHC~\cite{Acharya:2014nyr}: (light) 't Hooft-Polyakov monopoles~\cite{tHooft-Polyakov,Vento2013jua}, electroweak monopoles~\cite{Cho1996qd,cho2,You}, global monopoles~\cite{vilenkin,debate,nussinov,papav,sarkar} and monopolium~\cite{Diracs_idea,khlopov,Monopolium,Monopolium1}. Magnetic monopoles that carry a non-zero magnetic charge and dyons possessing both magnetic and electric charge are among the most fascinating  hypothetical particles. Even though there is no generally acknowledged empirical  evidence for their existence, there are strong theoretical reasons to believe that they do exist, and they are predicted by many theories including grand unified theories and superstring theory~\cite{Rajantie:2012xh,rajantiept}. 
 
%%%%

The theoretical motivation behind the introduction of magnetic monopoles is the symmetrisation of the Maxwell's equations and the explanation of the charge quantisation~\cite{Dirac1931kp}. Dirac showed that the mere existence of a monopole in the universe could offer an explanation of the discrete nature of the electric charge, leading to the Dirac Quantisation Condition (DQC),

\begin{equation} \alpha~  g = \frac{N}{2} e , \quad  N = 1, 2, ... , 
\label{eq:dqc}\end{equation}
 
\noindent where $e$ is the electron charge, $\alpha = \frac{e^2}{4\pi \hbar\, c \varepsilon_0 } = \frac{1}{137}$ is the fine structure constant (at zero energy, as appropriate to the fact that the quantisation condition of Dirac pertains to long (infrared) distances from the centre of the monopole), $\varepsilon_0$ is the vacuum permittivity, and $g$ is the monopole magnetic charge. In Dirac's formulation, magnetic monopoles are assumed to exist as point-like particles and quantum mechanical consistency conditions lead to Eq.~(\ref{eq:dqc}), establishing the value of their magnetic charge. Although monopoles symmetrise Maxwell's equations in form, there is a numerical asymmetry arising from the DQC, namely that the basic magnetic charge is much larger than the smallest electric charge. A magnetic monopole with a single Dirac charge ($g_{\rm D}$) has an equivalent electric charge of  $\beta(137e/2)$. Thus  for a relativistic monopole the energy loss is around $4,\!700$ times ($68.5^2$) that of a minimum-ionising electrically-charged particle. The monopole mass remains a free parameter of the theory.

A possible explanation for the lack of experimental confirmation of monopoles is Dirac's proposal~\cite{Dirac1931kp,Diracs_idea,khlopov} that monopoles are not seen freely because they form a bound state called \emph{monopolium}~\cite{Monopolium,Monopolium1,Epele0} being confined by strong magnetic forces. Monopolium is a neutral state, hence it is difficult to detect directly at a collider detector, although its decay into two photons would give a rather clear signal for the ATLAS and CMS detectors~\cite{Epele1,Epele2}, which however would not be visible in the MoEDAL detector. Nevertheless according to a novel proposal~\cite{risto}, the LHC radiation detector systems can be used to turn the LHC itself into a new physics search machine by detecting final-state protons $pp\to pXp$ exiting the LHC beam vacuum chamber at locations determined by their fractional momentum losses. Such technique would be appealing for detecting monopolia. Furthermore the monopolium might break up in the medium of MoEDAL into highly-ionising dyons, which subsequently can be detected in MoEDAL~\cite{Acharya:2014nyr}. Moreover its decay via photon emission would produce a peculiar trajectory in the medium, should the decaying states are also magnetic multipoles~\cite{Acharya:2014nyr}.

%%%%%%%%%%%%%%%%%%%%%%%%%%%%%%%%%%%%%%%%%%%%%%%%%%%%%%%%%%%%%%%%%%%%%%%%%%%%%%%%%%%%%%%%%%%%%%%%%%%%%%
\section{Searches for monopoles in MoEDAL}\label{sc:lightsearch}

The high ionisation of slow-moving magnetic monopoles and dyons, implies quite characteristic trajectories when such particles interact with the MoEDAL NTDs, which can be revealed during the etching process~\cite{moedal-tdr,Acharya:2014nyr}. In addition, the high magnetic charge of a monopole (which is expected to be at least one Dirac charge $g_D = 68.5 e$ (\emph{cf.} Eq.~(\ref{eq:dqc})) implies a strong magnetic dipole moment, which in turn may result in a strong binding of the monopole with the $^{27}_{13}{\rm Al}$ nuclei of the aluminium MoEDAL MMTs. In such a case, the presence of a monopole trapped in an aluminium bar of an MMT would de detected through the existence of a persistent current, defined as the difference between the currents in the SQUID of a magnetometer before and after the passage of the bar through the sensing coil. 

\begin{figure}[ht]
  \includegraphics[width=0.505\textwidth]{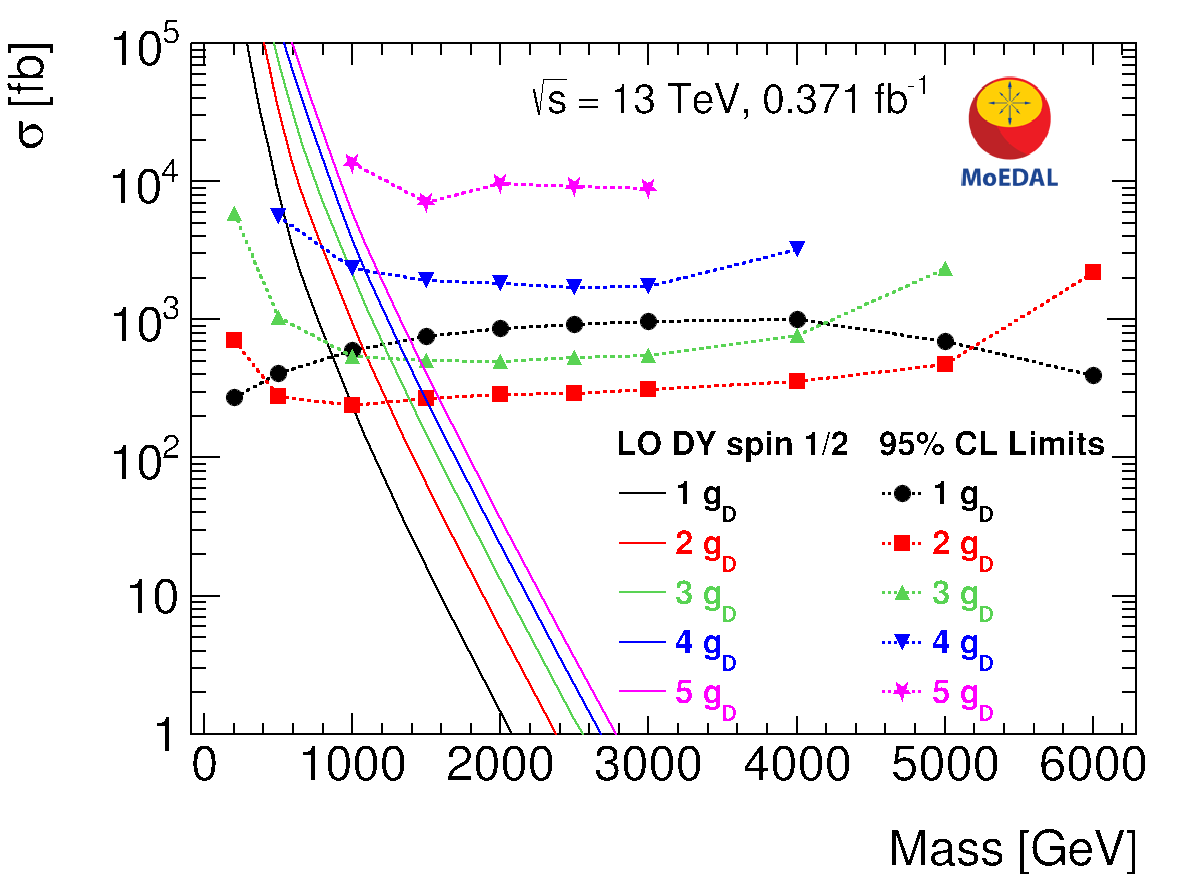}
  \includegraphics[width=0.505\textwidth]{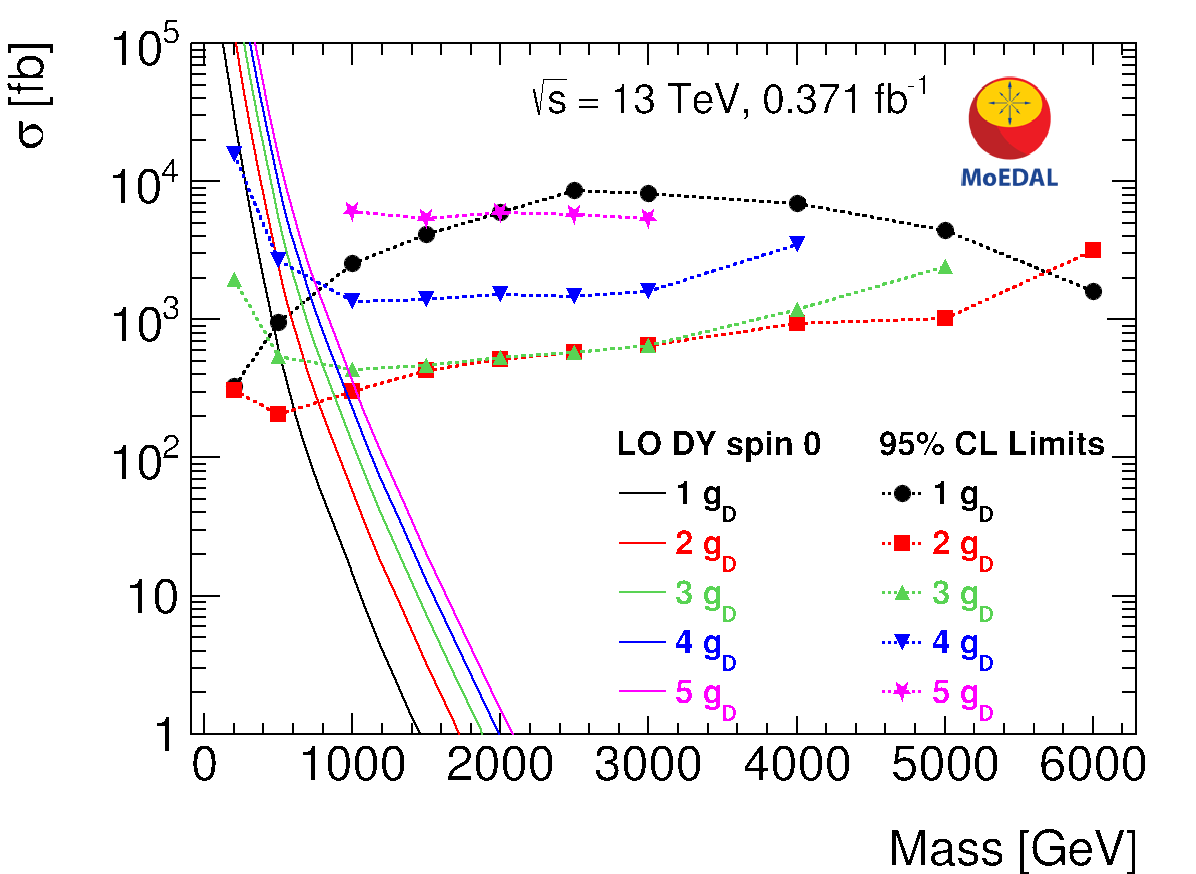}
  \caption{Cross-section upper limits at 95\% confidence level for DY monopole production as a function of mass for spin-1/2 (left) and for spin-0 monopoles (right). The various line styles correspond to different monopole charges. The solid lines represent DY cross-section calculations at leading order. From Ref.~\cite{MMT13TeV}.}
\label{fig:cross_section_limits}
\end{figure}

In the context of MoEDAL searches, two configurations of MMTs have been used at two different LHC c.m.\ energies as described in Section~\ref{sc:mmt}. No magnetic charge exceeding $0.5g_{\rm D}$ was detected in any of the exposed samples when passed through the ETH Zurich SQUID facility, allowing limits to be placed on monopole production. Model-independent cross-section limits have been obtained in fiducial regions of monopole energy and direction for $1g_{\rm D}\leq|g|\leq 6g_{\rm D}$ with the 8-TeV analysis~\cite{MMT8TeV}. Model-dependent cross-section limits are obtained for Drell-Yan (DY) pair production of spin-1/2 and spin-0 monopoles for $1g_{\rm D}\leq|g|\leq 5g_{\rm D}$ at 13~TeV~\cite{MMT13TeV}, as shown in Fig.~\ref{fig:cross_section_limits}. Caution, however, should be exerted here in the sense that the non-perturbative nature of the large magnetic Dirac charge of the monopole invalidate any perturbative treatment based on Drell-Yan calculations of the pertinent cross sections and hence any result based on the latter is only indicative, due to the lack of any other concrete theoretical treatment.

The weaker limits for $|g|= g_{\rm D}$ displayed in Fig.~\ref{fig:cross_section_limits} when compared to higher charges are mostly due to loss of acceptance from monopoles punching through the trapping volume. For higher charges, monopoles ranging out before reaching the trapping volume decrease the acceptance for DY monopoles with increasing charge and reaches below 0.1\% for a charge of $6g_{\rm D}$. The spin dependence is solely due to the different event kinematics: more central and more energetic monopoles for spin 0. 

\begin{figure}[ht]
  \includegraphics[width=0.66\linewidth]{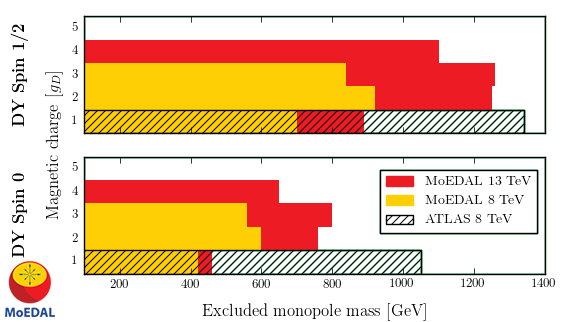}
  \caption{Excluded monopole masses for DY production for spin-$1/2$ (top) and spin-$0$ (bottom) monopoles. The MoEDAL results obtained at 8~TeV (yellow, light grey)~\cite{MMT8TeV} and 13~TeV (red, dark grey)~\cite{MMT13TeV} are superimposed on the ATLAS 8-TeV limits (hatched area)~\cite{atlas8tev}.}
\label{fig:exclusion3}
\end{figure}

Under the assumption of Drell-Yan cross sections, mass limits are derived for $g_{\rm D}\leq|g|\leq4g_{\rm D}$ at the LHC, complementing previous results from ATLAS Collaboration~\cite{atlas7tev,atlas8tev}, which placed limits for monopoles with magnetic charge $|g|\leq1.5 g_{\rm D}$ (c.f.\ Fig.~\ref{fig:exclusion3}). The ATLAS bounds are better that the MoEDAL ones for $|g|=1 g_{\rm D}$ due to the higher luminosity delivered in ATLAS and the loss of acceptance in MoEDAL for small magnetic charges. On the other hand, higher charges are difficult to be probed in ATLAS due to the limitations of the electromagnetic-calorimeter-based level-1 trigger deployed for such searches. A comparison of the limits on monopole production cross sections set by other colliders with those set by MoEDAL is presented in Ref.~\cite{rajantiept}, while general limits including searches in cosmic radiation are reviewed in Ref.~\cite{patrizii}. 

%%%%%%%%%%%%%%%%%%%%%%%%%%%%%%%%%%%%%%%%%%%%%%%%%%%
%%%%%%%%%%%%%%%%%%%%%%%%%%%%%%%%%%%%%%%%%%%%%%%%%%%
\section{Beyond magnetic monopoles}\label{sc:physics}

%%%%%%%%%%%%%%%%%%%%%%%%%%%%%%%%%%%%%%%%%%%%%%%%%%%
\subsection{Electrically-charged long-lived particles in supersymmetry}\label{sc:susy}

Supersymmetry (SUSY) is an extension of the Standard Model which assigns to each SM field a superpartner field with a spin differing by a half unit. SUSY provides elegant solutions to several open issues in the SM, such as the hierarchy problem, the identity of dark matter, and the grand unification. SUSY scenarios propose a number of massive slowly moving electrically charged particles. If they  are sufficiently long-lived to  travel a distance of at least ${\cal O}(1{\rm m})$  before decaying and their $Z/\beta\gtrsim 0.5$,  then they will be detected in the MoEDAL NTDs. No highly-charged particles are expected in such a theory, but there are several scenarios in which supersymmetry may yield massive, long-lived particles that could have electric charges $\pm 1$, potentially detectable in MoEDAL if they are produced with low velocities ($\beta \lesssim 0.2$)  .

The lightest supersymmetric particle (LSP) is stable in models where $R$~parity is conserved. The LSP should have no strong or electromagnetic interactions, for otherwise it would bind to conventional matter and be detectable in anomalous heavy nuclei~\cite{EHNOS}. Possible weakly-interacting neutral candidates in the Minimal Supersymmetric Standard Model (MSSM) include the sneutrino, which has been excluded by LEP and direct searches, the lightest neutralino $\tilde{\chi}_1^0$ (a mixture of spartners of the $Z, H$ and $\gamma$) and the gravitino $\tilde{G}$.

%%%%%%%%%%%%%%%%%%%%%%%%%%%%%%%%%%%%%%%%%%%%%%%%%%%
\subsubsection{Supersymmetric scenarios with $R$-parity violation}

Several scenarios featuring metastable charged sparticles might be detectable in MoEDAL. One such scenario is that $R$~parity {\it may not be exact}, since there is no exact local symmetry associated with either $L$ or $B$, and hence no fundamental reason why they should be conserved. One could consider various ways in which $L$ and/or $B$ could be violated in such a way that $R$ is violated, as represented by the following superpotential terms:
\begin{equation}
W_{RV} \; = \; \lambda^{\prime \prime}_{ijk} {\bar U}_i {\bar D}_j {\bar D}_k
+  \lambda^{\prime}_{ijk} {L}_i {Q}_j {\bar D}_k
+ \lambda_{ijk} {L}_i {L}_j {\bar E}_k
+ \mu_i L_i H,
\label{Rviolation}
\end{equation}
where ${Q}_i, {\bar U}_i, {\bar D}_i, L_i$ and ${\bar E}_i$ denote chiral superfields corresponding to quark doublets, antiquarks, lepton doublets and antileptons, respectively, with $i, j, k$ generation indices. The simultaneous presence of terms of the first and third type in Eq.~(\ref{Rviolation}), namely $\lambda$ and $\lambda^{\prime \prime}$, is severely restricted by lower limits on the proton lifetime, but other combinations are less restricted. The trilinear couplings in Eq.~(\ref{Rviolation}) generate sparticle decays such as ${\tilde q} \to {\bar q} {\bar q}$ or $q \ell$, or ${\tilde \ell} \to \ell \ell$, whereas the bilinear couplings in Eq.~(\ref{Rviolation}) generate Higgs-slepton mixing and thereby also ${\tilde q} \to q \ell$ and ${\tilde \ell} \to \ell \ell$ decays~\cite{Mitsou:2015kpa}. For a nominal sparticle mass $\sim 1$~TeV, the lifetime for such decays would exceed a few nanoseconds for $\lambda,  \lambda^{\prime}, \lambda^{\prime \prime} < 10^{-8}$. 

If $R$~parity is broken, the LSP would be unstable, and might be charged and/or coloured. In the former case, it might be detectable directly at the LHC as a massive slowly-moving charged particle. In the latter case, the LSP would bind with light quarks and/or gluons to make colour-singlet states, the so-called \emph{R-hadrons}, and any charged state could again be detectable as a massive slowly-moving charged particle. If $\lambda \ne 0$, the prospective experimental signature would be similar to a stau next-to-lightest sparticle (NLSP) case to be discussed later. On the other hand, if $\lambda^{\prime}$ or $\lambda^{\prime \prime} \ne 0$, the prospective experimental signature would be similar to a stop NLSP case, yielding the possibility of charge-changing interactions while passing through matter. This could yield  a metastable charged particle, created whilst passing through the material surrounding the intersection point,  that would be detected by MoEDAL. 

%%%%%%%%%%%%%%%%%%%%%%%%%%%%%%%%%%%%%%%%%%%%%%%%%%%
\subsubsection{Metastable lepton NLSP in the CMSSM with a neutralino LSP}

However, even if $R$~parity {\it is} exact, the NLSP may be long lived. This would occur, for example, if the LSP is the gravitino, or if the mass difference between the NLSP and the neutralino LSP is small, offering more scenarios for long-lived charged sparticles. In {\it neutralino dark matter} scenarios based on the constrained MSSM (CMSSM), for instance, the most natural candidate for the NLSP is the lighter stau slepton ${\tilde \tau_1}$~\cite{stauNLSP}, which could be long lived if $m_{\tilde \tau_1} - m_{\tilde{\chi}_1^0}$ is small. There are several regions of the CMSSM parameter space that are compatible with the constraints imposed by unsuccessful searches for sparticles at the LHC, as well as the discovered Higgs boson mass. These include a strip in the focus-point region where the relic density of the LSP is brought down into the range allowed by cosmology because of its relatively large Higgsino component, a region where the relic density is controlled by rapid annihilation through direct-channel heavy Higgs resonances, and a strip where the relic LSP density is reduced by coannihilations with near-degenerate staus and other sleptons. It was found in a global analysis that the two latter possibilities are favoured~\cite{MC8}.

\begin{figure}[ht]
\includegraphics[width=0.5\textwidth]{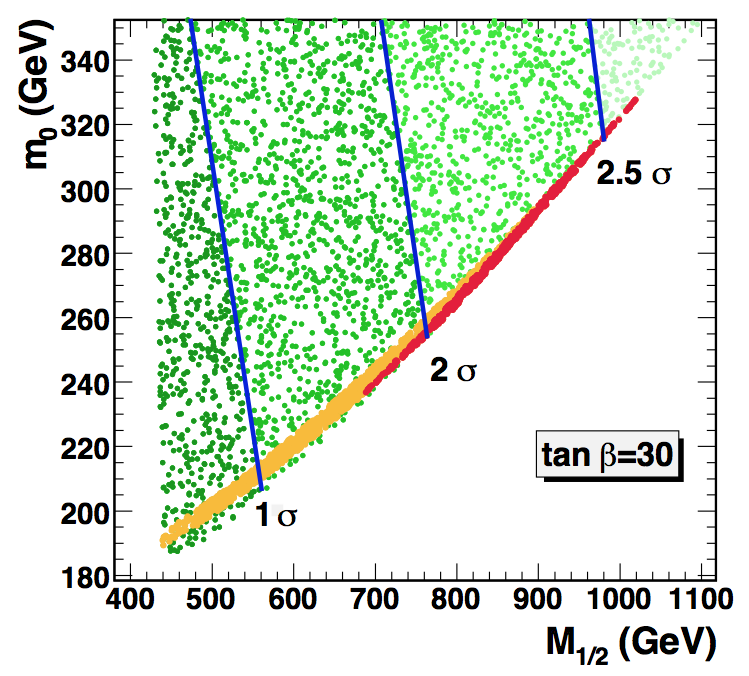}\hspace{2pc}%
\begin{minipage}[b]{0.43\textwidth}\caption{\label{fg:oscar}Allowed parameter regions in $M_{1/2} - m_0$ plane fixing $A_0=600$~GeV and $\tan\beta=30$. The red (dark) narrow band is consistent region with dark matter abundance and $\delta m < m_\tau$ and the yellow (light) narrow band is that with $\delta m > m_\tau$. The green regions are inconsistent with the dark matter abundance, and in the white (excluded) area the LSP is the stau. The favoured regions of the muon anomalous magnetic moment at $1\sigma$, $2\sigma$ and $2.5\sigma$ confidence level are indicated by solid lines. From Ref.~\cite{oscar}.}
\end{minipage}
\end{figure}

In the coannihilation region of the CMSSM, the lighter ${\tilde \tau_1}$ is expected to be the lightest slepton~\cite{stauNLSP}, and the $\tilde\tau_1-\tilde{\chi}_1^0$ mass difference may well be smaller than $m_\tau$: indeed, this is required at large LSP masses. In this case, the dominant stau decays for $m_{\tilde \tau_1} - m_{\tilde{\chi}_1^0} > 160$~MeV are expected to be into three particles: $\tilde{\chi}_1^0 \nu \pi$ or $\tilde{\chi}_1^0 \nu \rho$. If $m_{\tilde \tau_1} - m_{\tilde{\chi}_1^0} < 1.2$~GeV, the ${\tilde \tau_1}$ lifetime is calculated to be so long, in excess of $\sim 100$~ns, that it is likely to escape the detector before decaying, and hence would be detectable as a massive, slowly-moving charged particle~\cite{Sato,oscar}. The relevance of such scenarios while considering cosmological constraints is demonstrated in Fig.~\ref{fg:oscar}. Even is lepton-flavour violating couplings are allowed, the long lifetime of the staus remains~\cite{oscar}.

%%%%%%%%%%%%%%%%%%%%%%%%%%%%%%%%%%%%%%%%%%%%%%%%%%%
\subsubsection{Metastable sleptons in gravitino LSP scenarios}

On the other hand, in {\it gravitino dark matter} scenarios with more general options for the pattern of supersymmetry breaking, other options appear quite naturally, including the lighter selectron or smuon, or a sneutrino~\cite{sleptonNLSP}, or the lighter stop squark ${\tilde t_1}$~\cite{stopNLSP}. If the gravitino ${\tilde G}$ is the LSP, the decay rate of a slepton NLSP is given by 
\begin{equation}
\Gamma ( {\tilde \ell} \to {\tilde G} \ell) = \dfrac{1}{48 \pi M_*^2} \dfrac{m_{\tilde \ell}^5}{M_{\tilde G}^2}
\left[ 1 - \dfrac{M_{\tilde G}^2}{m_{\tilde \ell}^2} \right]^{4},
\label{telldecay}
\end{equation}
where $M_*$ is the Planck scale. Since $M_*$ is much larger than the electroweak scale, the NLSP lifetime is naturally very long.   

Gravitino (or axino) LSP with a long-lived charged stau may arise in gauge mediation and minimal supergravity models~\cite{Nojiri}. Large part of the parameter space potentially attractive for long-lived slepton searches with MoEDAL are compatible with cosmological constraints on the dark-matter abundance in superweakly interacting massive particle scenarios~\cite{Feng}.

%%%%%%%%%%%%%%%%%%%%%%%%%%%%%%%%%%%%%%%%%%%%%%%%%%%
\subsubsection{Long-lived gluinos in split supersymmetry}

The above discussion has been in the context of the CMSSM and similar scenarios where all the supersymmetric partners of Standard Model particles have masses in the TeV range. Another scenario is ``split supersymmetry'', in which the supersymmetric partners of quarks and leptons are very heavy, of a scale $m_s$, whilst the supersymmetric partners of SM bosons are relatively light~\cite{splitSUSY}. In such a case, the gluino could have a mass in the TeV range and hence be accessible to the LHC, but would have a very long lifetime:
\begin{equation}
\tau \approx 8 \left( \dfrac{m_s}{10^9~{\rm GeV}} \right)^4 \left( \dfrac{1~{\rm TeV}}{m_{\tilde{g}}} \right)^5~{\rm s}.
\label{gluinotau}
\end{equation}
Long-lived gluinos would form long-lived gluino R-hadrons including gluino-gluon \emph{(gluinoball)} combinations, gluino-$q{\bar q}$ \emph{(mesino)} combinations and gluino-$qqq$ \emph{(baryino)} combinations. The heavier gluino hadrons would be expected to decay into the lightest species, which would be metastable, with a lifetime given by Eq.~(\ref{gluinotau}), and it is possible that this metastable gluino hadron could be charged.

In the same way as stop hadrons, gluino hadrons may flip charge through conventional strong interactions as they pass through matter, and it is possible that one may pass through most of a conventional LHC tracking detector undetected in a neutral state before converting into a metastable charged state that could be detected by MoEDAL. 

%%%%%%%%%%%%%%%%%%%%%%%%%%%%%%%%%%%%%%%%%%%%%%%%%%%
\subsubsection{Experimental considerations} 

There are several considerations supporting the complementary aspects of MoEDAL w.r.t.\ ATLAS and CMS when discussing the observability of (meta-)stable massive electrically-charged particles. Most of them stem from MoEDAL being ``time-agnostic'' due to the passive nature of its detectors. Therefore signal from very slowly moving particles will not be lost due to arriving in several consecutive bunch crossings. ATLAS and CMS, on the other hand, perform triggered-based analyses relying either on triggering on accompanying ``objects'', e.g.\ missing transverse energy, or by developing and deploying specialised triggers. In both cases the efficiency may lower and in the former the probed parameter space may be reduced. MoEDAL is mainly limited by the geometrical acceptance of the detectors, especially the MMTs, and by the requirement of passing the $Z/\beta$ threshold of NTDs. In general ATLAS and CMS have demonstrated to cover high-velocities $\beta \gtrsim 0.2$, while MoEDAL is sensitive to lower ones $\beta \lesssim 0.2$. 

When discussing the detection of particles stopped \emph{(trapped)} in material that they may decay later, different possibilities are explored. CMS and ATLAS look in empty bunch crossings for decays of trapped particles into jets. MoEDAL MMTs may be monitored in a underground/basement laboratory for tracks arising from such decays. The background in the latter case, coming from cosmic rays, should be easier to control and assess. The probed lifetimes should be larger due to the unlimited monitoring time. 

%%%%%%%%%%%%%%%%%%%%%%%%%%%%%%%%%%%%%%%%%%%%%%%%%%%
\subsection{Doubly-charged Higgs bosons}\label{sc:lrsm} 

Doubly-charged particles appear in many theoretical scenarios beyond the SM. For example, doubly-charged scalar states, usually termed doubly-charged Higgs fields, appear in left-right symmetric models~ \cite{Pati1974yy,LRSM,LRSMa} and in see-saw models for neutrino masses with Higgs triplets. A number of models encompasses additional symmetries and extend the SM Higgs sector by introducing doubly-charged Higgs bosons. A representative example of such a model is the L-R Symmetric Model (LRSM)~\cite{Pati1974yy,LRSM,LRSMa}, proposed to amend the fact that the weak-interaction couplings are strictly left handed by extending the gauge group of the SM so as to include a right-handed sector. The simplest realisation is an LRSM~\cite{Pati1974yy, LRSM}  that postulates  a right-handed version of the weak interaction, whose gauge symmetry is spontaneously broken at high mass scale, leading to the parity-violating  SM. This model naturally accommodates recent data on neutrino oscillations and the existence of small neutrino masses. The model generally requires Higgs triplets containing doubly-charged Higgs bosons ($H^{\pm\pm}$)  $\Delta_{R}^{++}$ and $\Delta_{L}^{++}$, which could be light in the minimal supersymmetric left-right model~\cite{LRSUSY}.

Single production of a doubly-charged Higgs boson at the LHC proceeds via vector boson fusion, or through the fusion of a singly-charged Higgs boson with either a $W^\pm$ or another singly-charged Higgs boson. The amplitudes of the  $W_{L} W_{L}$ and $W_{R} W_{R}$ vector boson fusion processes are proportional to $v_{L,R}$, the vacuum expectation values of the neutral members of the scalar triplets of the  LRSM. For the case of $\Delta_{R}^{++}$ production, the vector boson fusion process dominates. Pair production of doubly-charged Higgs bosons is also possible via a Drell-Yan process, with $\gamma$, $Z$ or $Z_{R}$ exchanged in the $s$-channel, but at a high kinematic price since substantial energy is required to produce two heavy particles. In the case of $\Delta_{L}^{++}$, double production may nevertheless be the only possibility if $v_{L}$ is very small or vanishing.

\begin{figure}[ht]
\includegraphics[width=0.5\textwidth]{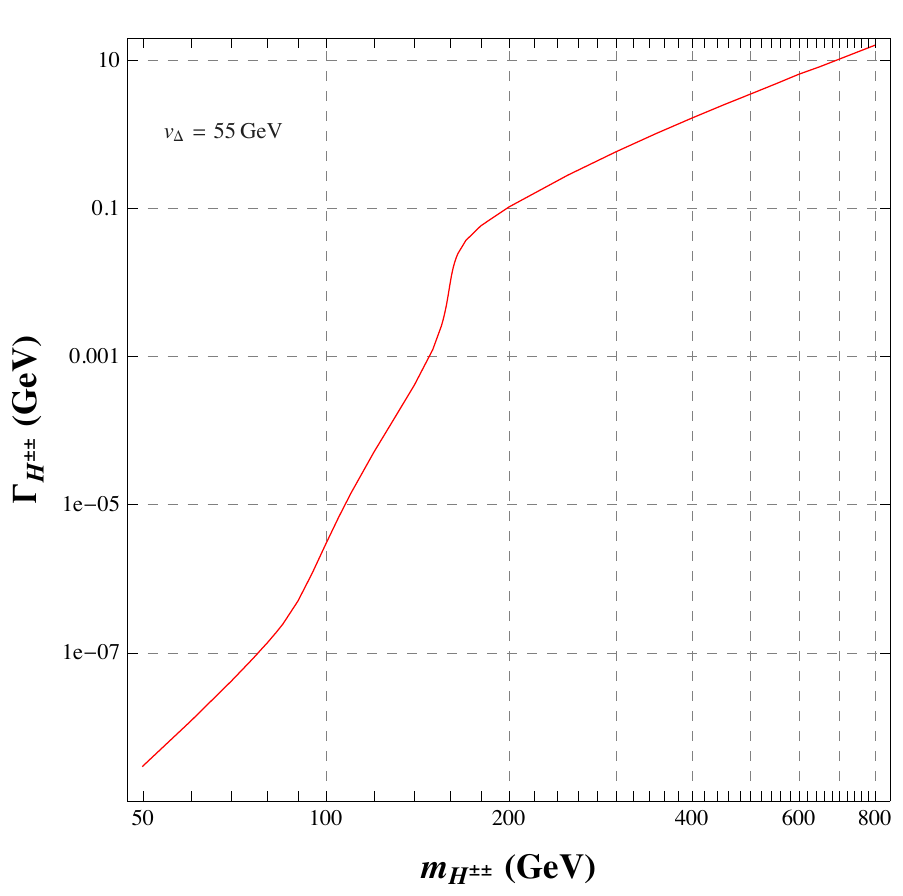}\hspace{2pc}%
\begin{minipage}[b]{0.43\textwidth}\caption{\label{fig:width}Partial decay width of $H^{\pm \pm} \rightarrow W^{\pm} W^{\pm}$ as a function of $m_{H^{\pm\pm}}$ for $v_{\Delta} = 55~{\rm GeV}$. From Ref.~\cite{Chiang:2012dk}.}
\end{minipage}
\end{figure}

The decay of a doubly-charged Higgs boson can proceed via several channels. The dilepton signature leads to the (experimentally clean) final state $q\bar{q} \rightarrow \Delta^{++}_L\Delta^{--}_L \rightarrow 4\ell$. However as long as the triplet vacuum expectation value, $v_\Delta$, is much larger than $10^{-4}~{\rm GeV}$, the doubly-charged Higgs decay predominantly into a pair of same-sign $W$ bosons. For very small Yukawa couplings $H_{\ell\ell} \lesssim  10^{-8}$, the doubly-charged Higgs boson can be quasi-stable~\cite{Chiang:2012dk}. In Fig.~\ref{fig:width}, the partial decay width of the doubly charged Higgs boson into a $W$ boson pair is shown as a function of its mass. For $v_\Delta \gg 10^{-4}~{\rm GeV}$, this partial width is roughly equal to the total width of the doubly charged Higgs boson. In the case of long lifetimes, slowly moving pseudo-stable Higgs bosons could be detected in the MoEDAL NTDs. For example with CR39, one could detect doubly-charged Higgs particles moving with speeds less than around $\beta \simeq 0.4$. 

%%%%%%%%%%%%%%%%%%%%%%%%%%%%%%%%%%%%%%%%%%%%%%%%%%%
\subsection{Black hole remnants in large extra dimensions}\label{sc:bh}

Over the last decades, models based on compactified extra spatial dimensions (ED) have been proposed in order to explain the large gap between the electroweak (EW) and the Planck scale of $M_{\rm EW}/M_{\rm PL}  \approx 10^{-17}$. The four main scenarios relevant for searches at LHC the Arkani-Hamed-Dimopoulos-Dvali (ADD) model of large extra dimensions~\cite{ADD}, the Randall-Sundrum (RS) model of warped extra dimensions~\cite{Randall}, TeV$^{-1}$-sized extra dimensions~\cite{TEV-1}, and the Universal Extra Dimensions (UED) model~\cite{UED}.

The existence of extra spatial dimensions~\cite{ADD,Randall} and a sufficiently small fundamental scale of gravity open up the possibility that microscopic black holes be produced and detected~\cite{ED1, bhevaporation, fischler1, CHARYBDIS, ED2,ED3}  at the LHC. Once produced, the black holes will undergo an evaporation process categorised in three stages~\cite{bhevaporation, fischler1}: the \emph{balding phase}, the actual \emph{evaporation phase}, and finally   the Planck phase. It is generally assumed that the black hole will decay completely to some last few SM particles. However, another intriguing possibility is that the remaining energy is carried away by a stable remnant.

The prospect of microscopic black hole production at the LHC within the framework of models with large extra dimensions has been studied in Ref.~\cite{ADD}. Black holes produced at the LHC are expected to decay with an average multiplicity of $\sim10-25$ into SM particles,  most of which will be charged, though the details of the multiplicity distribution depend on the number of extra dimensions~\cite{BHMULT}. After the black holes have evaporated off enough energy to reach the remnant mass, some will have accumulated a net electric charge. According to purely statistical considerations, the probability for being left with highly-charged black hole remnants drops fast with the deviation from the average. The largest fraction of the black holes should have charges $\pm1$ or zero, although a smaller but non-negligible fraction would be multiply charged.

The fraction of charged black-hole remnants has been estimated~\cite{BHMULT,Hossenfelder:2005ku} using the {\tt PYTHIA} event generator~\cite{PYTHIA} and the {\tt CHARYBDIS} program~\cite{CHARYBDIS}. It was  assumed that the effective temperature of the black hole drops towards zero for a finite remnant mass, $M_{R}$. The value of $M_{R}$ does not noticeably affect the investigated charge distribution, as it results from the very general statistical distribution of the charge of the emitted particles.

\begin{figure}[ht]
\includegraphics[width=0.5\textwidth]{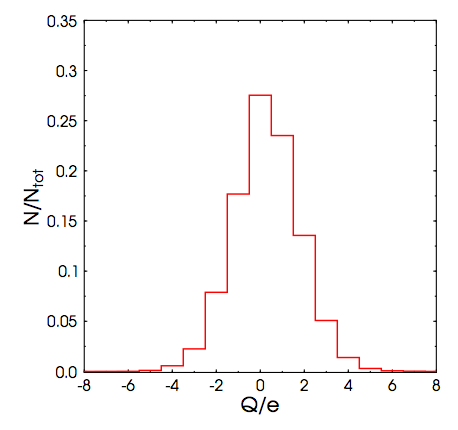}\hspace{2pc}%
\begin{minipage}[b]{0.43\textwidth}\caption{\label{Fig:Qofremnants}The distribution of black-hole remnant charges in proton-proton interactions at $\sqrt{s} = 14~{\rm TeV}$ calculated with the
{\tt PYTHIA} event generator~\cite{PYTHIA} and the {\tt CHARYBDIS} program~\cite{CHARYBDIS}. From Ref.~\cite{Hossenfelder:2005ku}.}
\end{minipage}
\end{figure}

Thus, independent of the underlying quantum-gravitational  assumption leading to the remnant formation, it was found that about 30\% of the remnants are neutral, whereas $\sim$ 40\% would be singly-charged black holes, and the  remaining $\sim$30\% of remnants would be multiply-charged.  The distribution of the  remnant charges obtained  is shown in Fig.~\ref{Fig:Qofremnants}. The black hole remnants  considered here are heavy, with masses of a TeV or more. A significant  fraction of the black-hole remnants produced would have a  Z/$\beta$ of greater than five, high enough to register in the CR39 NTDs forming the LT-NTD  sub-detector of MoEDAL.

%%%%%%%%%%%%%%%%%%%%%%%%%%%%%%%%%%%%%%%%%%%%%%%%%%%
\subsection{D-matter}\label{sc:dmatter}

Some versions of string theory include higher-dimensional ``domain-wall''-like membrane \emph{(brane)} structures in space-time, called \emph{D-branes}. In some cases the bulk is ``punctured'' by lower-dimensional D-brane defects, which are either point-like or have their longitudinal dimensions compactified~\cite{westmuckett}. From a low-energy observer's perspective, such structures would effectively appear to be point-like \emph{D-particles}. The latter have  dynamical degrees of freedom, thus they can be treated as quantum  excitations above the vacuum~\cite{westmuckett,shiu} collectively referred to as {\it D-matter}. D-matter states are non-perturbative stringy objects with masses of order $m_D \sim M_s/g_s$, where $g_s $ is the string coupling, typically of order one so that the observed gauge and gravitational couplings is reproduced. Hence, the D-matter states could be light enough to be phenomenologically relevant at the LHC.

Depending on their type, D-branes could carry integral or torsion (discrete) charges with the lightest D-particle (LDP) being stable. Therefore the LDPs are possible candidates for cold dark matter~\cite{shiu}. D-particles are solitonic non-perturbative objects in the string/brane theory. As discussed in the relevant literature~\cite{shiu}, there are similarities and differences between D-particles and magnetic monopoles with non-trivial cosmological implications~\cite{Witten2002wb,westmuckett,Mavromatos:2010jt,mitsou}. An important difference is that they could have {\it perturbative} couplings, with no magnetic charge in general. Nonetheless, in the context of brane-inspired gauge theories, brane states with magnetic charges can be constructed, which would manifest themselves in MoEDAL in a manner similar to  magnetic monopoles. 
 
\begin{figure}[ht]
\centering
\includegraphics[width=0.3\textwidth]{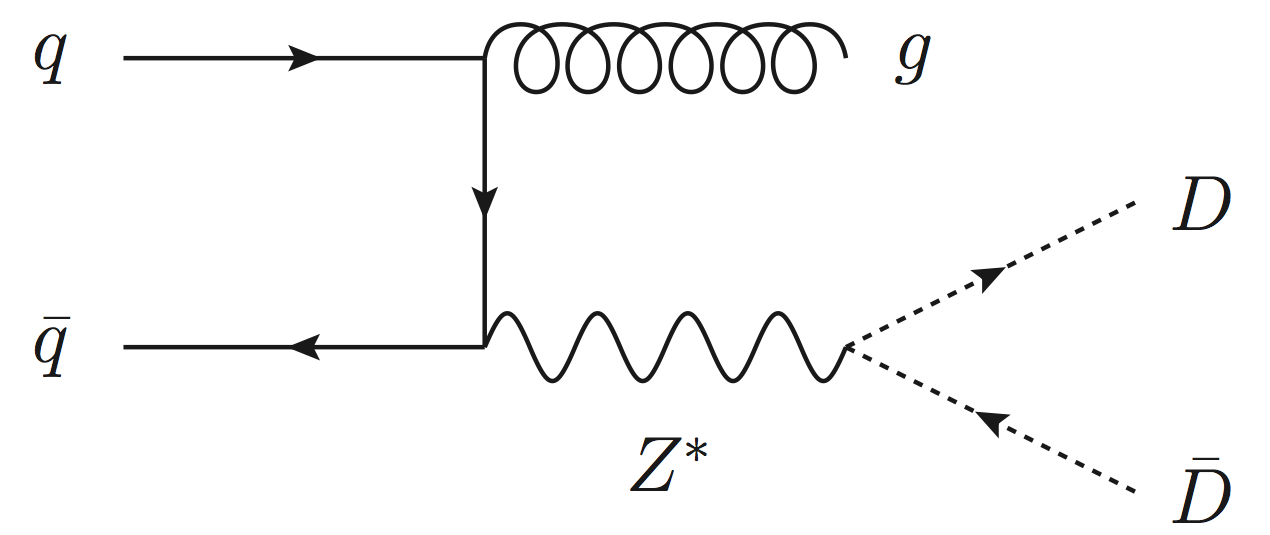}
\caption{An example of parton-level diagrams for production of D-particles by $q\bar{q}$ collisions in a generic D-matter low-energy model~\cite{mitsou}. }
\label{fig:dproduction}
\end{figure}

Non-magnetically-charged D-matter, on the other hand, could be produced at colliders and also produce interesting signals of direct relevance to the MoEDAL experiment. For instance, excited states of D-matter (${\rm D}^\star$) can be electrically-charged. For typical string couplings of phenomenological relevance, the first few massive levels may be accessible to the LHC.  Depending on the details of the microscopic model considered, and the way the SM is embedded, such massive charged states can be relatively long-lived, and could likewise be detectable with MoEDAL. D-matter/antimatter pairs can be produced~\cite{Mavromatos:2010jt,mitsou} by the decay of intermediate off-shell $Z$-bosons, as shown in Fig.~\ref{fig:dproduction}. 

%%%%%%%%%%%%%%%%%%%%%%%%%%%%%%%%%%%%%%%%%%%%%%%%%%%%%%%%%%%%%%%%%%%%%%%%%%%%%%%%%%%%%%%%%%%%%%%%%%%%%%
\section{Summary and outlook}\label{sc:summary}

MoEDAL is going to extend considerably the LHC reach in the search for (meta)stable highly ionising particles. The latter are predicted in a variety of theoretical models and include: magnetic monopoles, SUSY stable (or, rather, long-lived) spartners, quirks, strangelets, Q-balls, fractionally-charged massive particles, etc~\cite{Acharya:2014nyr}. Such particles can be light enough to be producible at the LHC energies (see e.g.\ Q-balls in the context of some SUSY or brane models~\cite{kehagias}). In this paper we have described searches for only a subset of those particles, due to lack of space. Specifically, we discussed monopoles, partners in some SUSY models, doubly charged Higgs bosons, black hole remnants in models with large extra spatial dimensions, as well as some (more exotic) scenarios on D-matter which characterises some brane models. 

The MoEDAL design is optimised to probe precisely all such long lived states, unlike the other LHC experiments~\cite{DeRoeck:2011aa}. Furthermore it combines different detector technologies: plastic nuclear track detectors (NTDs), trapping volumes and pixel sensors~\cite{moedal-tdr}. The first physics results, pertaining to magnetic monopole trapping detectors, obtained with LHC Run~1 data~\cite{MMT8TeV}, and the corresponding analysis at 13~TeV has been published recently~\cite{MMT13TeV}. The MoEDAL Collaboration is preparing new analyses with more Run~2 data, with other detectors (NTDs) and with a large variety of interpretations involving not only magnetic but also electric charges.

%%%%%%%%%%%%%%%%%%%%%%%%%%%%%%%%%%%%%%%%%%%%%%%%%%%
%%%%%%%%%%%%%%%%%%%%%%%%%%%%%%%%%%%%%%%%%%%%%%%%%%%
\section*{Acknowledgments}

The author is grateful to the DISCRETE2016 Symposium organisers for the kind invitation to present this talk. She acknowledges support by the Spanish Ministry of Economy and Competitiveness (MINECO) under the project FPA2015-65652-C4-1-R, by the Generalitat Valenciana through the project PROMETEO~II/2013-017, by the Spanish National Research Council (CSIC) under the CT Incorporation Program 201650I002 and by the Severo Ochoa Excellence Centre Project SEV 2014-0398.

%%%%%%%%%%%%%%%%%%%%%%%%%%%%%%%%%%%%%%%%%%%%%%%%%%%
%%%%%%%%%%%%%%%%%%%%%%%%%%%%%%%%%%%%%%%%%%%%%%%%%%%


\section*{References}
\begin{thebibliography}{99}

\bibitem{moedal-web} For general information on the MoEDAL experiment, see: \url{http://moedal.web.cern.ch/}

\bibitem{moedal-tdr} MoEDAL Collaboration, Technical Design Report of the MoEDAL Experiment {\it CERN Preprint} CERN-LHC-2009-006, MoEDAL-TDR-1.1 (2009), and references therein.

\bibitem{LHC}  L.~Evans and P.~Bryant,
  %``LHC Machine,''
  JINST {\bf 3}, S08001 (2008). 
  %doi:10.1088/1748-0221/3/08/S08001
  %%CITATION = doi:10.1088/1748-0221/3/08/S08001;%%

\bibitem{DeRoeck:2011aa}  A.~De Roeck, A.~Katre, P.~Mermod, D.~Milstead and T.~Sloan,
  %``Sensitivity of LHC Experiments to Exotic Highly Ionising Particles,''
  Eur.\ Phys.\ J.\ C {\bf 72}, 1985 (2012).
  %doi:10.1140/epjc/s10052-012-1985-2 [arXiv:1112.2999 [hep-ph]].
  %%CITATION = doi:10.1140/epjc/s10052-012-1985-2;%%  
  
\bibitem{Fairbairn07} M.~Fairbairn, A.~C.~Kraan, D.~A.~Milstead, T.~Sjostrand, P.~Z.~Skands and T.~Sloan,
  %``Stable massive particles at colliders,''
  Phys.\ Rept.\  {\bf 438}, 1 (2007);
  %doi:10.1016/j.physrep.2006.10.002 [hep-ph/0611040].
  %%CITATION = doi:10.1016/j.physrep.2006.10.002;%%
S.~Burdin, M.~Fairbairn, P.~Mermod, D.~Milstead, J.~Pinfold, T.~Sloan and W.~Taylor,
  %``Non-collider searches for stable massive particles,''
  Phys.\ Rept.\  {\bf 582}, 1 (2015).
  %doi:10.1016/j.physrep.2015.03.004 [arXiv:1410.1374 [hep-ph]].
  %%CITATION = doi:10.1016/j.physrep.2015.03.004;%%  
  
\bibitem{creta2016}
  V.~A.~Mitsou [MoEDAL Collaboration],
  %``The physics case for the MoEDAL experiment at LHC,''
  EPJ Web Conf.\  {\bf 95}, 04042 (2015);
  %doi:10.1051/epjconf/20159504042 [arXiv:1411.7651 [hep-ph]].
  %%CITATION = doi:10.1051/epjconf/20159504042;%%  
%  V.~A.~Mitsou [MoEDAL Collaboration],
  %``MoEDAL: Seeking magnetic monopoles and more at the LHC,''
  PoS {\bf EPS-HEP2015}, 109 (2015);   % [arXiv:1511.01745 [physics.ins-det]].
  %%CITATION = ARXIV:1511.01745;%%
  N.~E.~Mavromatos and V.~A.~Mitsou [MoEDAL Collaboration],
  ``Physics reach of MoEDAL at LHC: magnetic monopoles, supersymmetry and beyond,''
  EPJ Web Conf.\ {\it to appear} [arXiv:1612.07012 [hep-ph]].
  %%CITATION = ARXIV:1612.07012;%%
  
\bibitem{Acharya:2014nyr}  B.~Acharya {\it et al.} [MoEDAL Collaboration],
  %``The Physics Programme Of The MoEDAL Experiment At The LHC,''
  Int.\ J.\ Mod.\ Phys.\ A {\bf 29}, 1430050 (2014).
  %doi:10.1142/S0217751X14300506 [arXiv:1405.7662 [hep-ph]].
  %%CITATION = doi:10.1142/S0217751X14300506;%%
  
\bibitem{LHCb-detector}    A.~A.~Alves, Jr. {\it et al.} [LHCb Collaboration],
  %``The LHCb Detector at the LHC,''
  JINST {\bf 3}, S08005 (2008).
  %doi:10.1088/1748-0221/3/08/S08005
  %%CITATION = doi:10.1088/1748-0221/3/08/S08005;%%
  
\bibitem{Joergensen:2012gy}    M.~D.~Joergensen, A.~De Roeck, H.-P.~Hachler, A.~Hirt, A.~Katre, P.~Mermod, D.~Milstead and T.~Sloan,
  ``Searching for magnetic monopoles trapped in accelerator material at the Large Hadron Collider,''
  arXiv:1206.6793 [physics.ins-det] (2012).
  %%CITATION = ARXIV:1206.6793;%%  
  
\bibitem{DeRoeck:2012wua}  A.~De Roeck, H.~P.~H\"achler, A.~M.~Hirt, M.-D.~Joergensen, A.~Katre, P.~Mermod, D.~Milstead and T.~Sloan,
  %``Development of a magnetometer-based search strategy for stopped monopoles at the Large Hadron Collider,''
  Eur.\ Phys.\ J.\ C {\bf 72}, 2212 (2012).
  %doi:10.1140/epjc/s10052-012-2212-x
  %%CITATION = doi:10.1140/epjc/s10052-012-2212-x;%%
  
\bibitem{timepix}  A.~Sopczak {\it et al.},
  %``MPX Detectors as LHC Luminosity Monitor,''
  IEEE Trans.\ Nucl.\ Sci.\  {\bf 62},  3225 (2015).
  %doi:10.1109/TNS.2015.2496316  [arXiv:1512.08014 [physics.ins-det]].
  %%CITATION = doi:10.1109/TNS.2015.2496316;%%
    
\bibitem{Dirac1931kp} P.~A.~M.~Dirac,
  %``Quantized Singularities in the Electromagnetic Field,''
  Proc.\ Roy.\ Soc.\ Lond.\ A {\bf 133}, 60 (1931).
  %doi:10.1098/rspa.1931.0130
  %%CITATION = doi:10.1098/rspa.1931.0130;%%

\bibitem{Diracs_idea} P.~A.~M.~Dirac,
  %``The Theory of magnetic poles,''
  Phys.\ Rev.\  {\bf 74}, 817 (1948).
  %doi:10.1103/PhysRev.74.817
  %%CITATION = doi:10.1103/PhysRev.74.817;%%
  
\bibitem{tHooft-Polyakov} G.~'t Hooft,
  %``Magnetic Monopoles in Unified Gauge Theories,''
  Nucl.\ Phys.\ B {\bf 79}, 276 (1974);
  %doi:10.1016/0550-3213(74)90486-6
  %%CITATION = doi:10.1016/0550-3213(74)90486-6;%%%\bibitem{Polyakov1974ek} 
A.~M.~Polyakov,
  %``Particle Spectrum in the Quantum Field Theory,''
  JETP Lett.\  {\bf 20}, 194 (1974)
  [Pisma Zh.\ Eksp.\ Teor.\ Fiz.\  {\bf 20}, 430 (1974)].
  %%CITATION = JTPLA,20,194;%%
  
\bibitem{mm}  B.~Julia and A.~Zee,
  %``Poles with Both Magnetic and Electric Charges in Nonabelian Gauge Theory,''
  Phys.\ Rev.\ D {\bf 11}, 2227 (1975);  %doi:10.1103/PhysRevD.11.2227
  %%CITATION = doi:10.1103/PhysRevD.11.2227;%%
Y.~Nambu,
  %``String-Like Configurations in the Weinberg-Salam Theory,''
  Nucl.\ Phys.\ B {\bf 130}, 505 (1977);  %doi:10.1016/0550-3213(77)90252-8
  %%CITATION = doi:10.1016/0550-3213(77)90252-8;%%
E.~Witten,
  %``Dyons of Charge e theta/2 pi,''
  Phys.\ Lett.\  {\bf 86B}, 283 (1979);  %doi:10.1016/0370-2693(79)90838-4
  %%CITATION = doi:10.1016/0370-2693(79)90838-4;%%
G.~Lazarides, M.~Magg and Q.~Shafi,
  %``Phase Transitions and Magnetic Monopoles in SO(10),''
  Phys.\ Lett.\  {\bf 97B}, 87 (1980);  %doi:10.1016/0370-2693(80)90553-5
  %%CITATION = doi:10.1016/0370-2693(80)90553-5;%%
R.~d.~Sorkin,
  %``Kaluza-Klein Monopole,''
  Phys.\ Rev.\ Lett.\  {\bf 51}, 87 (1983);  %doi:10.1103/PhysRevLett.51.87
  %%CITATION = doi:10.1103/PhysRevLett.51.87;%%
D.~J.~Gross and M.~J.~Perry,
  %``Magnetic Monopoles in Kaluza-Klein Theories,''
  Nucl.\ Phys.\ B {\bf 226}, 29 (1983);  %doi:10.1016/0550-3213(83)90462-5
  %%CITATION = doi:10.1016/0550-3213(83)90462-5;%%
J.~S.~Schwinger,
  %``A Magnetic model of matter,''
  Science {\bf 165}, 757 (1969);  %doi:10.1126/science.165.3895.757
  %%CITATION = doi:10.1126/science.165.3895.757;%% 
J.~Preskill,
  %``Magnetic Monopoles,''
  Ann.\ Rev.\ Nucl.\ Part.\ Sci.\  {\bf 34}, 461 (1984);  %doi:10.1146/annurev.ns.34.120184.002333
  %%CITATION = doi:10.1146/annurev.ns.34.120184.002333;%% 
A.~Achucarro and T.~Vachaspati,
  %``Semilocal and electroweak strings,''
  Phys.\ Rept.\  {\bf 327}, 347 (2000)  [Phys.\ Rept.\  {\bf 327}, 427 (2000)];
  %doi:10.1016/S0370-1573(99)00103-9  [hep-ph/9904229].
  %%CITATION = doi:10.1016/S0370-1573(99)00103-9;%%
T.~W.~Kephart, C.~A.~Lee and Q.~Shafi,
  %``Family unification, exotic states and light magnetic monopoles,''
  JHEP {\bf 0701}, 088 (2007); %doi:10.1088/1126-6708/2007/01/088   [hep-ph/0602055].
  %%CITATION = doi:10.1088/1126-6708/2007/01/088;%%
D.~G.~Pak, P.~M.~Zhang and L.~P.~Zou,
  %``On finite energy monopole solutions in Weinberg?Salam model,''
  Int.\ J.\ Mod.\ Phys.\ A {\bf 30}, no. 27, 1550164 (2015); %doi:10.1142/S0217751X1550164X [arXiv:1311.7567 [hep-th]].
  %%CITATION = doi:10.1142/S0217751X1550164X;%%%\bibitem{Rajantie2005hi} 
A.~Rajantie,
  %``Mass of a quantum 't Hooft-Polyakov monopole,''
  JHEP {\bf 0601}, 088 (2006).  % doi:10.1088/1126-6708/2006/01/088   [hep-lat/0512006].
  %%CITATION = doi:10.1088/1126-6708/2006/01/088;%%
  
 \bibitem{Vento2013jua}  V.~Vento and V.~S.~Mantovani,
 ``On the magnetic monopole mass,''
  arXiv:1306.4213 [hep-ph] (2013).

\bibitem{Cho1996qd}  Y.~M.~Cho and D.~Maison,
  %``Monopoles in Weinberg-Salam model,''
  Phys.\ Lett.\ B {\bf 391}, 360 (1997); % doi:10.1016/S0370-2693(96)01492-X  [hep-th/9601028].
  %%CITATION = doi:10.1016/S0370-2693(96)01492-X;%%
W.~S.~Bae and Y.~M.~Cho,
  %``Finite energy electroweak dyon,''
  J.\ Korean Phys.\ Soc.\  {\bf 46}, 791 (2005).  %[hep-th/0210299].
  %%CITATION = HEP-TH/0210299;%%
  
 \bibitem{cho2} Y.~M.~Cho, K.~Kimm and J.~H.~Yoon,
  %``Mass of the Electroweak Monopole,''
  Mod.\ Phys.\ Lett.\ A {\bf 31}, no. 09, 1650053 (2016); % doi:10.1142/S021773231650053X [arXiv:1212.3885 [hep-ph]].
  %%CITATION = doi:10.1142/S021773231650053X;%%
 %Y.~M.~Cho, K.~Kimm and J.~H.~Yoon,
  %``Gravitationally Coupled Electroweak Monopole,''
  Phys.\ Lett.\ B {\bf 761}, 203 (2016).  %  doi:10.1016/j.physletb.2016.08.033  [arXiv:1605.08129 [hep-th]].
  %%CITATION = doi:10.1016/j.physletb.2016.08.033;%% 
 
 \bibitem{You} J.~Ellis, N.~E.~Mavromatos and T.~You,
  %``The Price of an Electroweak Monopole,''
  Phys.\ Lett.\ B {\bf 756}, 29 (2016);  % doi:10.1016/j.physletb.2016.02.048 [arXiv:1602.01745 [hep-ph]].
  %%CITATION = doi:10.1016/j.physletb.2016.02.048;%%
J.~Ellis, V.~Sanz and T.~You,
  %``Complete Higgs Sector Constraints on Dimension-6 Operators,''
  JHEP {\bf 1407}, 036 (2014). %  doi:10.1007/JHEP07(2014)036  [arXiv:1404.3667 [hep-ph]].
  %%CITATION = doi:10.1007/JHEP07(2014)036;%%

 \bibitem{vilenkin} M.~Barriola and A.~Vilenkin,
  %``Gravitational Field of a Global Monopole,''
  Phys.\ Rev.\ Lett.\  {\bf 63}, 341 (1989).  %  doi:10.1103/PhysRevLett.63.341
  %%CITATION = doi:10.1103/PhysRevLett.63.341;%%

\bibitem{debate} A.~S.~Goldhaber,
  %``Collapse of a 'Global Monopole.',''
  Phys.\ Rev.\ Lett.\  {\bf 63}, 2158 (1989); %  doi:10.1103/PhysRevLett.63.2158;
  %%CITATION = doi:10.1103/PhysRevLett.63.2158;%%
S.~H.~Rhie and D.~P.~Bennett,
  %``Global monopoles do not 'collapse',''
  Phys.\ Rev.\ Lett.\  {\bf 67}, 1173 (1991);  % doi:10.1103/PhysRevLett.67.1173;
  %%CITATION = doi:10.1103/PhysRevLett.67.1173;%%
L.~Perivolaropoulos,
  %``Instabilities and interactions of global topological defects,''
  Nucl.\ Phys.\ B {\bf 375}, 665 (1992);  % doi:10.1016/0550-3213(92)90115-R;
  %%CITATION = doi:10.1016/0550-3213(92)90115-R;%%
G.~W.~Gibbons, M.~E.~Ortiz, F.~Ruiz Ruiz and T.~M.~Samols,
  %``Semilocal strings and monopoles,''
  Nucl.\ Phys.\ B {\bf 385}, 127 (1992); % doi:10.1016/0550-3213(92)90097-U  [hep-th/9203023];
  %%CITATION = doi:10.1016/0550-3213(92)90097-U;%%
M.~Hindmarsh,
  %``Semilocal topological defects,''
  Nucl.\ Phys.\ B {\bf 392}, 461 (1993); % doi:10.1016/0550-3213(93)90681-E  [hep-ph/9206229];
  %%CITATION = doi:10.1016/0550-3213(93)90681-E;%%
G.~Arreaga, I.~Cho and J.~Guven,
  %``Stability of selfgravitating magnetic monopoles,''
  Phys.\ Rev.\ D {\bf 62}, 043520 (2000); % doi:10.1103/PhysRevD.62.043520  [gr-qc/0001078];
  %%CITATION = doi:10.1103/PhysRevD.62.043520;%%
A.~Achucarro and J.~Urrestilla,
  %``The (In)stability of global monopoles revisited,''
  Phys.\ Rev.\ Lett.\  {\bf 85}, 3091 (2000); % doi:10.1103/PhysRevLett.85.3091  [hep-ph/0003145].
  %%CITATION = doi:10.1103/PhysRevLett.85.3091;%%
R.~Gregory and C.~Santos,
  %``Space-time structure of the global vortex,''
  Class.\ Quant.\ Grav.\  {\bf 20}, 21 (2003);  % doi:10.1088/0264-9381/20/1/302  [hep-th/0208037].
  %%CITATION = doi:10.1088/0264-9381/20/1/302;%%
S.~B.~Gudnason and J.~Evslin,
  %``Global monopoles of charge 2,''
  Phys.\ Rev.\ D {\bf 92}, no. 4, 045044 (2015); % doi:10.1103/PhysRevD.92.045044 [arXiv:1507.03400 [hep-th]];
  %%CITATION = doi:10.1103/PhysRevD.92.045044;%%
A.~Achucarro, B.~Hartmann and J.~Urrestilla,
  %``Exotic composites: The Decay of deficit angles in global-local monopoles,''
  JHEP {\bf 0507}, 006 (2005).  % doi:10.1088/1126-6708/2005/07/006  [hep-th/0504192];
  %%CITATION = doi:10.1088/1126-6708/2005/07/006;%%

\bibitem{nussinov} A.~K.~Drukier and S.~Nussinov,
  %``Monopole Pair Creation in Energetic Collisions: Is It Possible?,''
  Phys.\ Rev.\ Lett.\  {\bf 49}, 102 (1982). % doi:10.1103/PhysRevLett.49.102
  %%CITATION = doi:10.1103/PhysRevLett.49.102;%%

\bibitem{papav} P.~O.~Mazur and J.~Papavassiliou,
  %``Gravitational scattering on a global monopole,''
  Phys.\ Rev.\ D {\bf 44}, 1317 (1991).  %  doi:10.1103/PhysRevD.44.1317
  %%CITATION = doi:10.1103/PhysRevD.44.1317;%%

\bibitem{sarkar} 
  N.~E.~Mavromatos and S.~Sarkar,
  ``Magnetic Monopoles from Global Monopoles in the presence of Kalb-Ramond Torsion,''
  arXiv:1607.01315 [hep-th] (2016).
  %%CITATION = ARXIV:1607.01315;%% 

\bibitem{khlopov} Y.~B.~Zeldovich and M.~Y.~Khlopov,
  %``On the Concentration of Relic Magnetic Monopoles in the Universe,''
  Phys.\ Lett.\  {\bf 79B}, 239 (1978).  % doi:10.1016/0370-2693(78)90232-0
  %%CITATION = doi:10.1016/0370-2693(78)90232-0;%%

\bibitem{Monopolium} C.~T.~Hill,
  %``Monopolonium,''
  Nucl.\ Phys.\ B {\bf 224}, 469 (1983).  %  doi:10.1016/0550-3213(83)90386-3
  %%CITATION = doi:10.1016/0550-3213(83)90386-3;%%

\bibitem{Monopolium1} V.~K.~Dubrovich,
  %``Association of magnetic monopoles and antimonopoles in the early universe,''
  Grav.\ Cosmol.\ Suppl.\  {\bf 8N1}, 122 (2002).
  %%CITATION = 00292,8N1,122;%%

\bibitem{Rajantie:2012xh}   A.~Rajantie,
  %``Introduction to Magnetic Monopoles,''
  Contemp.\ Phys.\  {\bf 53}, 195 (2012). %doi:10.1080/00107514.2012.685693  [arXiv:1204.3077 [hep-th]].
  %%CITATION = doi:10.1080/00107514.2012.685693;%% 

\bibitem{rajantiept}  A.~Rajantie,
  %``The search for magnetic monopoles,''
  Phys.\ Today {\bf 69}, no. 10, 40 (2016).
  %doi:10.1063/PT.3.3328
  %%CITATION = doi:10.1063/PT.3.3328;%%

\bibitem{Epele0}  L.~N.~Epele, H.~Fanchiotti, C.~A.~Garcia Canal and V.~Vento,
  %``Monopolium: The Key to monopoles,''
  Eur.\ Phys.\ J.\ C {\bf 56}, 87 (2008);  % doi:10.1140/epjc/s10052-008-0628-0  [hep-ph/0701133].
  %%CITATION = doi:10.1140/epjc/s10052-008-0628-0;%% 
%L.~N.~Epele, H.~Fanchiotti, C.~A.~Garcia Canal and V.~Vento, 
L.~N.~Epele, H.~Fanchiotti, C.~A.~G.~Canal and V.~Vento,
  %``Monopolium production from photon fusion at the Large Hadron Collider,''
  Eur.\ Phys.\ J.\ C {\bf 62}, 587 (2009).   % doi:10.1140/epjc/s10052-009-1069-0  [arXiv:0809.0272 [hep-ph]].
  %%CITATION = doi:10.1140/epjc/s10052-009-1069-0;%%    

\bibitem{Epele1} L.~N.~Epele, H.~Fanchiotti, C.~A.~G.~Canal, V.~A.~Mitsou and V.~Vento,
  %``Looking for magnetic monopoles at LHC with diphoton events,''
  Eur.\ Phys.\ J.\ Plus {\bf 127}, 60 (2012).  % doi:10.1140/epjp/i2012-12060-8  [arXiv:1205.6120 [hep-ph]].
  %%CITATION = doi:10.1140/epjp/i2012-12060-8;%%

\bibitem{Epele2}  L.~N.~Epele, H.~Fanchiotti, C.~A.~G.~Canal, V.~A.~Mitsou and V.~Vento,
  ``Can the 750~GeV enhancement be a signal of light magnetic monopoles?,''
  arXiv:1607.05592 [hep-ph] (2016).
  %%CITATION = ARXIV:1607.05592;%%  

\bibitem{risto}  M.~Kalliokoski, J.~W.~L\"ams\"a, M.~Mieskolainen and R.~Orava,
  ``Turning the LHC Ring into a New Physics Search Machine,''
  arXiv:1604.05778 [hep-ex] (2016). 
  %%CITATION = ARXIV:1604.05778;%%   

\bibitem{MMT8TeV} B.~Acharya {\it et al.} [MoEDAL Collaboration],
  %``Search for magnetic monopoles with the MoEDAL prototype trapping detector in 8 TeV proton-proton collisions at the LHC,''
  JHEP {\bf 1608}, 067 (2016).  % doi:10.1007/JHEP08(2016)067  [arXiv:1604.06645 [hep-ex]].
  %%CITATION = doi:10.1007/JHEP08(2016)067;%%

\bibitem{MMT13TeV} B.~Acharya {\it et al.} [MoEDAL Collaboration],
  %``Search for magnetic monopoles with the MoEDAL forward trapping detector in 13 TeV proton-proton collisions at the LHC,''
  Phys.\ Rev.\ Lett.\  {\bf 118},  061801 (2017).
  %doi:10.1103/PhysRevLett.118.061801 [arXiv:1611.06817 [hep-ex]].
  %%CITATION = doi:10.1103/PhysRevLett.118.061801;%%
  
\bibitem{atlas7tev}  G.~Aad {\it et al.} [ATLAS Collaboration],
  %``Search for magnetic monopoles in $\sqrt{s}=7$ TeV $pp$ collisions with the ATLAS detector,''
  Phys.\ Rev.\ Lett.\  {\bf 109}, 261803 (2012).  % doi:10.1103/PhysRevLett.109.261803  [arXiv:1207.6411 [hep-ex]];
  %%CITATION = doi:10.1103/PhysRevLett.109.261803;%%
  
\bibitem{atlas8tev}   G.~Aad {\it et al.} [ATLAS Collaboration],
  %``Search for magnetic monopoles and stable particles with high electric charges in 8 TeV $pp$ collisions with the ATLAS detector,''
  Phys.\ Rev.\ D {\bf 93}, 052009 (2016).   % doi:10.1103/PhysRevD.93.052009 [arXiv:1509.08059 [hep-ex]].
  %%CITATION = doi:10.1103/PhysRevD.93.052009;%%

 \bibitem{patrizii}  L.~Patrizii and M.~Spurio,
  %``Status of Searches for Magnetic Monopoles,''
  Ann.\ Rev.\ Nucl.\ Part.\ Sci.\  {\bf 65}, 279 (2015).
%  doi:10.1146/annurev-nucl-102014-022137
%  [arXiv:1510.07125 [hep-ex]].
  %%CITATION = doi:10.1146/annurev-nucl-102014-022137;%% 
  
\bibitem{EHNOS}H.~Goldberg,
  %``Constraint on the Photino Mass from Cosmology,''
  Phys.\ Rev.\ Lett.\  {\bf 50}, 1419 (1983)
  Erratum: [Phys.\ Rev.\ Lett.\  {\bf 103}, 099905 (2009)];   % doi:10.1103/PhysRevLett.50.1419
  %%CITATION = doi:10.1103/PhysRevLett.50.1419;%%
J.~R.~Ellis, J.~S.~Hagelin, D.~V.~Nanopoulos, K.~A.~Olive and M.~Srednicki,
  %``Supersymmetric Relics from the Big Bang,''
  Nucl.\ Phys.\ B {\bf 238}, 453 (1984).  %   doi:10.1016/0550-3213(84)90461-9
  %%CITATION = doi:10.1016/0550-3213(84)90461-9;%%

\bibitem{Mitsou:2015kpa} V.~A.~Mitsou,
  %``R-parity violating supersymmetry and neutrino physics: experimental signatures,''
  PoS {\bf PLANCK~2015}, 085 (2015).
  %[arXiv:1510.02660 [hep-ph]].
  %%CITATION = ARXIV:1510.02660;%%
      
\bibitem{stauNLSP}  J.~R.~Ellis, K.~A.~Olive, Y.~Santoso and V.~C.~Spanos,
  %``Supersymmetric dark matter in light of WMAP,''
  Phys.\ Lett.\ B {\bf 565}, 176 (2003).  % doi:10.1016/S0370-2693(03)00765-2  [hep-ph/0303043].
  %%CITATION = doi:10.1016/S0370-2693(03)00765-2;%%

\bibitem{MC8} O.~Buchmueller {\it et al.},
  %``The CMSSM and NUHM1 in Light of 7 TeV LHC, $B_s \to \mu^+\mu^-$ and XENON100 Data,''
  Eur.\ Phys.\ J.\ C {\bf 72}, 2243 (2012);  % doi:10.1140/epjc/s10052-012-2243-3 [arXiv:1207.7315 [hep-ph]].
  %%CITATION = doi:10.1140/epjc/s10052-012-2243-3;%% 
%O.~Buchmueller {\it et al.},
  %``The CMSSM and NUHM1 after LHC Run 1,''
  Eur.\ Phys.\ J.\ C {\bf 74}, 2922 (2014).  % doi:10.1140/epjc/s10052-014-2922-3 [arXiv:1312.5250 [hep-ph]].
  %%CITATION = doi:10.1140/epjc/s10052-014-2922-3;%%    
  
\bibitem{Sato}  T.~Jittoh, J.~Sato, T.~Shimomura and M.~Yamanaka,
  %``Long life stau in the minimal supersymmetric standard model,''
  Phys.\ Rev.\ D {\bf 73}, 055009 (2006)
  Erratum: [Phys.\ Rev.\ D {\bf 87}, 019901 (2013)].  
  % doi:10.1103/PhysRevD.73.055009, 10.1103/PhysRevD.87.019901 [hep-ph/0512197].
  %%CITATION = doi:10.1103/PhysRevD.73.055009, 10.1103/PhysRevD.87.019901;%%
    
\bibitem{oscar}  S.~Kaneko, J.~Sato, T.~Shimomura, O.~Vives and M.~Yamanaka,
  %``Measuring lepton flavor violation at LHC with a long-lived slepton in the coannihilation region,''
  Phys.\ Rev.\ D {\bf 78}, 116013 (2008)
  Erratum: [Phys.\ Rev.\ D {\bf 87}, 039904 (2013)].
  %doi:10.1103/PhysRevD.78.116013, 10.1103/PhysRevD.87.039904 [arXiv:0811.0703 [hep-ph]].
  %%CITATION = doi:10.1103/PhysRevD.78.116013, 10.1103/PhysRevD.87.039904;%%
  
\bibitem{sleptonNLSP} J.~R.~Ellis, K.~A.~Olive and Y.~Santoso,
  %``Sneutrino NLSP Scenarios in the NUHM with Gravitino Dark Matter,''
  JHEP {\bf 0810}, 005 (2008).   % doi:10.1088/1126-6708/2008/10/005 [arXiv:0807.3736 [hep-ph]].
  %%CITATION = doi:10.1088/1126-6708/2008/10/005;%%

\bibitem{stopNLSP}  J.~L.~Diaz-Cruz, J.~R.~Ellis, K.~A.~Olive and Y.~Santoso,
  %``On the Feasibility of a Stop NLSP in Gravitino Dark Matter Scenarios,''
  JHEP {\bf 0705}, 003 (2007).  % doi:10.1088/1126-6708/2007/05/003 [hep-ph/0701229 [HEP-PH]].
  %%CITATION = doi:10.1088/1126-6708/2007/05/003;%%
    
\bibitem{Nojiri} K.~Hamaguchi, M.~M.~Nojiri and A.~de Roeck,
  %``Prospects to study a long-lived charged next lightest supersymmetric particle at the LHC,''
  JHEP {\bf 0703}, 046 (2007).  % doi:10.1088/1126-6708/2007/03/046 [hep-ph/0612060].
  %%CITATION = doi:10.1088/1126-6708/2007/03/046;%%
  
\bibitem{Feng}  J.~L.~Feng, S.~Iwamoto, Y.~Shadmi and S.~Tarem,
  %``Long-Lived Sleptons at the LHC and a 100 TeV Proton Collider,''
  JHEP {\bf 1512}, 166 (2015).  % doi:10.1007/JHEP12(2015)166 [arXiv:1505.02996 [hep-ph]].
  %%CITATION = doi:10.1007/JHEP12(2015)166;%%
    
\bibitem{splitSUSY}  N.~Arkani-Hamed and S.~Dimopoulos,
  %``Supersymmetric unification without low energy supersymmetry and signatures for fine-tuning at the LHC,''
  JHEP {\bf 0506}, 073 (2005);   % doi:10.1088/1126-6708/2005/06/073 [hep-th/0405159].
  %%CITATION = doi:10.1088/1126-6708/2005/06/073;%%
G.~F.~Giudice and A.~Romanino,
  %``Split supersymmetry,''
  Nucl.\ Phys.\ B {\bf 699}, 65 (2004)
  Erratum: [Nucl.\ Phys.\ B {\bf 706}, 487 (2005)]. 
  %doi:10.1016/j.nuclphysb.2004.11.048, 10.1016/j.nuclphysb.2004.08.001 [hep-ph/0406088].
  %%CITATION = doi:10.1016/j.nuclphysb.2004.11.048, 10.1016/j.nuclphysb.2004.08.001;%%
  
\bibitem{Pati1974yy} J.~C.~Pati and A.~Salam,
  %``Lepton Number as the Fourth Color,''
  Phys.\ Rev.\ D {\bf 10}, 275 (1974)
  Erratum: [Phys.\ Rev.\ D {\bf 11}, 703 (1975)].  % doi:10.1103/PhysRevD.10.275, 10.1103/PhysRevD.11.703.2
  %%CITATION = doi:10.1103/PhysRevD.10.275, 10.1103/PhysRevD.11.703.2;%%
 
\bibitem{LRSM} R.~N.~Mohapatra and J.~C.~Pati,
  %``Left-Right Gauge Symmetry and an Isoconjugate Model of CP Violation,''
  Phys.\ Rev.\ D {\bf 11}, 566 (1975).  %  doi:10.1103/PhysRevD.11.566
  %%CITATION = doi:10.1103/PhysRevD.11.566;%%
  
\bibitem{LRSMa}   G.~Senjanovic and R.~N.~Mohapatra,
  %``Exact Left-Right Symmetry and Spontaneous Violation of Parity,''
  Phys.\ Rev.\ D {\bf 12}, 1502 (1975);  %  doi:10.1103/PhysRevD.12.1502
  %%CITATION = doi:10.1103/PhysRevD.12.1502;%%
R.~N.~Mohapatra and G.~Senjanovic,
  %``Neutrino Masses and Mixings in Gauge Models with Spontaneous Parity Violation,''
  Phys.\ Rev.\ D {\bf 23}, 165 (1981).  % doi:10.1103/PhysRevD.23.165
  %%CITATION = doi:10.1103/PhysRevD.23.165;%%

\bibitem{LRSUSY} C.~S.~Aulakh, A.~Melfo and G.~Senjanovic,
  %``Minimal supersymmetric left-right model,''
  Phys.\ Rev.\ D {\bf 57}, 4174 (1998);  % doi:10.1103/PhysRevD.57.4174  [hep-ph/9707256].
  %%CITATION = doi:10.1103/PhysRevD.57.4174;%%
Z.~Chacko and R.~N.~Mohapatra,
  %``Supersymmetric left-right model and light doubly charged Higgs bosons and Higgsinos,''
  Phys.\ Rev.\ D {\bf 58}, 015003 (1998);  % doi:10.1103/PhysRevD.58.015003 [hep-ph/9712359].
  %%CITATION = doi:10.1103/PhysRevD.58.015003;%%
%\bibitem{LRSUSY2} 
C.~S.~Aulakh, K.~Benakli and G.~Senjanovic,
  %``Reconciling supersymmetry and left-right symmetry,''
  Phys.\ Rev.\ Lett.\  {\bf 79}, 2188 (1997).  % doi:10.1103/PhysRevLett.79.2188 [hep-ph/9703434].
  %%CITATION = doi:10.1103/PhysRevLett.79.2188;%%
 
\bibitem{Chiang:2012dk} C.~W.~Chiang, T.~Nomura and K.~Tsumura,
  %``Search for doubly charged Higgs bosons using the same-sign diboson mode at the LHC,''
  Phys.\ Rev.\ D {\bf 85}, 095023 (2012).  % doi:10.1103/PhysRevD.85.095023 [arXiv:1202.2014 [hep-ph]].
  %%CITATION = doi:10.1103/PhysRevD.85.095023;%%

\bibitem{ADD}  N.~Arkani-Hamed, S.~Dimopoulos and G.~R.~Dvali,
  %``The Hierarchy problem and new dimensions at a millimeter,''
  Phys.\ Lett.\ B {\bf 429}, 263 (1998);  % doi:10.1016/S0370-2693(98)00466-3  [hep-ph/9803315].
  %%CITATION = doi:10.1016/S0370-2693(98)00466-3;%%
%N.~Arkani-Hamed, S.~Dimopoulos, G.~R.~Dvali, 
%``Phenomenology, astrophysics and cosmology of theories with submillimeter dimensions and TeV scale quantum gravity,''
  Phys.\ Rev.\ D {\bf 59}, 086004 (1999);  % doi:10.1103/PhysRevD.59.086004  [hep-ph/9807344].
  %%CITATION = doi:10.1103/PhysRevD.59.086004;%%
I.~Antoniadis, N.~Arkani-Hamed, S.~Dimopoulos and G.~R.~Dvali,
  %``New dimensions at a millimeter to a Fermi and superstrings at a TeV,''
  Phys.\ Lett.\ B {\bf 436}, 257 (1998).   % doi:10.1016/S0370-2693(98)00860-0  [hep-ph/9804398].
  %%CITATION = doi:10.1016/S0370-2693(98)00860-0;%%

\bibitem{Randall} L.~Randall and R.~Sundrum,
  %``An Alternative to compactification,''
  Phys.\ Rev.\ Lett.\  {\bf 83}, 4690 (1999).  % doi:10.1103/PhysRevLett.83.4690  [hep-th/9906064].
  %%CITATION = doi:10.1103/PhysRevLett.83.4690;%% 
  
\bibitem{TEV-1} I.~Antoniadis,
  %``A Possible new dimension at a few TeV,''
  Phys.\ Lett.\ B {\bf 246}, 377 (1990);  %  doi:10.1016/0370-2693(90)90617-F
  %%CITATION = doi:10.1016/0370-2693(90)90617-F;%% 
I.~Antoniadis and K.~Benakli,
  %``Limits on extra dimensions in orbifold compactifications of superstrings,''
  Phys.\ Lett.\ B {\bf 326}, 69 (1994);  % doi:10.1016/0370-2693(94)91194-0  [hep-th/9310151].
  %%CITATION = doi:10.1016/0370-2693(94)91194-0;%%
I.~Antoniadis, K.~Benakli and M.~Quiros,
  %``Production of Kaluza-Klein states at future colliders,''
  Phys.\ Lett.\ B {\bf 331}, 313 (1994).  % doi:10.1016/0370-2693(94)91058-8  [hep-ph/9403290].
  %%CITATION = doi:10.1016/0370-2693(94)91058-8;%%

\bibitem{UED} T.~Appelquist, H.~C.~Cheng and B.~A.~Dobrescu,
  %``Bounds on universal extra dimensions,''
  Phys.\ Rev.\ D {\bf 64}, 035002 (2001).  % doi:10.1103/PhysRevD.64.035002  [hep-ph/0012100].
  %%CITATION = doi:10.1103/PhysRevD.64.035002;%% 

\bibitem{bhevaporation} S.~B.~Giddings and S.~D.~Thomas,
  %``High-energy colliders as black hole factories: The End of short distance physics,''
  Phys.\ Rev.\ D {\bf 65}, 056010 (2002).  % doi:10.1103/PhysRevD.65.056010  [hep-ph/0106219].
  %%CITATION = doi:10.1103/PhysRevD.65.056010;%%

\bibitem{fischler1} W.~Fischler, 
  ``A Model for high-energy scattering in quantum gravity,'' 
  arXiv:hep-th/9906038 (1999).

\bibitem{ED1} P.~C.~Argyres, S.~Dimopoulos and J.~March-Russell,
  %``Black holes and submillimeter dimensions,''
  Phys.\ Lett.\ B {\bf 441}, 96 (1998);  % doi:10.1016/S0370-2693(98)01184-8  [hep-th/9808138].
  %%CITATION = doi:10.1016/S0370-2693(98)01184-8;%%
S.~Dimopoulos and G.~L.~Landsberg,
  %``Black holes at the LHC,''
  Phys.\ Rev.\ Lett.\  {\bf 87}, 161602 (2001);   % doi:10.1103/PhysRevLett.87.161602  [hep-ph/0106295].
  %%CITATION = doi:10.1103/PhysRevLett.87.161602;%%
G.~L.~Alberghi, R.~Casadio and A.~Tronconi,
  %``Quantum Gravity Effects in Black Holes at the LHC,''
  J.\ Phys.\ G {\bf 34}, 767 (2007);  % doi:10.1088/0954-3899/34/4/012 [hep-ph/0611009].
  %%CITATION = doi:10.1088/0954-3899/34/4/012;%%%\bibitem{ED1c} 
 M.~Cavaglia, R.~Godang, L.~Cremaldi and D.~Summers,
  %``Catfish: A Monte Carlo simulator for black holes at the LHC,''
  Comput.\ Phys.\ Commun.\  {\bf 177}, 506 (2007);  % doi:10.1016/j.cpc.2007.05.011 [hep-ph/0609001].
  %%CITATION = doi:10.1016/j.cpc.2007.05.011;%% 
D.~C.~Dai, G.~Starkman, D.~Stojkovic, C.~Issever, E.~Rizvi and J.~Tseng,
  %``BlackMax: A black-hole event generator with rotation, recoil, split branes, and brane tension,''
  Phys.\ Rev.\ D {\bf 77}, 076007 (2008).  % doi:10.1103/PhysRevD.77.076007 [arXiv:0711.3012 [hep-ph]].
  %%CITATION = doi:10.1103/PhysRevD.77.076007;%%
  
\bibitem{ED2} R.~Casadio and B.~Harms,
  %``Can black holes and naked singularities be detected in accelerators?,''
  Int.\ J.\ Mod.\ Phys.\ A {\bf 17}, 4635 (2002).  % doi:10.1142/S0217751X02012259  [hep-th/0110255].
  %%CITATION = doi:10.1142/S0217751X02012259;%%

\bibitem{ED3} M.~Cavaglia,
  %``Black hole and brane production in TeV gravity: A Review,''
  Int.\ J.\ Mod.\ Phys.\ A {\bf 18}, 1843 (2003);  % doi:10.1142/S0217751X03013569 [hep-ph/0210296].
  %%CITATION = doi:10.1142/S0217751X03013569;%%
P.~Kanti,
  %``Black holes in theories with large extra dimensions: A Review,''
  Int.\ J.\ Mod.\ Phys.\ A {\bf 19}, 4899 (2004).  % doi:10.1142/S0217751X04018324  [hep-ph/0402168].
  %%CITATION = doi:10.1142/S0217751X04018324;%%

\bibitem{CHARYBDIS} C.~M.~Harris, P.~Richardson and B.~R.~Webber,
  %``CHARYBDIS: A Black hole event generator,''
  JHEP {\bf 0308}, 033 (2003).  % doi:10.1088/1126-6708/2003/08/033  [hep-ph/0307305].
  %%CITATION = doi:10.1088/1126-6708/2003/08/033;%% 

\bibitem{BHMULT} B.~Koch, M.~Bleicher and H.~Stoecker,
  %``Black Holes at LHC?,''
  J.\ Phys.\ G {\bf 34}, S535 (2007).  % doi:10.1088/0954-3899/34/8/S44  [hep-ph/0702187 [HEP-PH]].
  %%CITATION = doi:10.1088/0954-3899/34/8/S44;%%

\bibitem{Hossenfelder:2005ku}  S.~Hossenfelder, B.~Koch and M.~Bleicher,
  ``Trapping black hole remnants,'' 
  hep-ph/0507140 (2005).
  %%CITATION = HEP-PH/0507140;%% 
  
\bibitem{PYTHIA} T.~Sjostrand, S.~Mrenna and P.~Z.~Skands,
  %``PYTHIA 6.4 Physics and Manual,''
  JHEP {\bf 0605}, 026 (2006).   % doi:10.1088/1126-6708/2006/05/026 [hep-ph/0603175].
  %%CITATION = doi:10.1088/1126-6708/2006/05/026;%%
  
\bibitem{westmuckett} J.~R.~Ellis, N.~E.~Mavromatos and D.~V.~Nanopoulos,
  %``Time dependent vacuum energy induced by D particle recoil,''
  Gen.\ Rel.\ Grav.\  {\bf 32}, 943 (2000);  % doi:10.1023/A:1001993226227  [gr-qc/9810086].
  %%CITATION = doi:10.1023/A:1001993226227;%% 
J.~R.~Ellis, N.~E.~Mavromatos and D.~V.~Nanopoulos,
  %``Derivation of a Vacuum Refractive Index in a Stringy Space-Time Foam Model,''
  Phys.\ Lett.\ B {\bf 665}, 412 (2008);  % doi:10.1016/j.physletb.2008.06.029  [arXiv:0804.3566 [hep-th]].
  %%CITATION = doi:10.1016/j.physletb.2008.06.029;%%
J.~R.~Ellis, N.~E.~Mavromatos and M.~Westmuckett,
  %``A Supersymmetric D-brane model of space-time foam,''
  Phys.\ Rev.\ D {\bf 70}, 044036 (2004);   % doi:10.1103/PhysRevD.70.044036  [gr-qc/0405066].
  %%CITATION = doi:10.1103/PhysRevD.70.044036;%%% \bibitem{westmuckett3} 
%J.~R.~Ellis, N.~E.~Mavromatos and M.~Westmuckett,
  %``Potentials between D-branes in a supersymmetric model of space-time foam,''
  {\it ibid.} {\bf 71}, 106006 (2005).  % doi:10.1103/PhysRevD.71.106006  [gr-qc/0501060].
  %%CITATION = doi:10.1103/PhysRevD.71.106006;%%
     
\bibitem{shiu}  G.~Shiu and L.~T.~Wang,
  %``D matter,''
  Phys.\ Rev.\ D {\bf 69}, 126007 (2004).   % doi:10.1103/PhysRevD.69.126007 [hep-ph/0311228].
  %%CITATION = doi:10.1103/PhysRevD.69.126007;%%

\bibitem{Mavromatos:2010jt} N.~E.~Mavromatos, S.~Sarkar and A.~Vergou,
  %``Stringy Space-Time Foam, Finsler-like Metrics and Dark Matter Relics,''
  Phys.\ Lett.\ B {\bf 696}, 300 (2011).   % doi:10.1016/j.physletb.2010.12.045  [arXiv:1009.2880 [hep-th]].
  %%CITATION = doi:10.1016/j.physletb.2010.12.045;%%
  
\bibitem{mitsou}  N.~E.~Mavromatos, V.~A.~Mitsou, S.~Sarkar and A.~Vergou,
  %``Implications of a Stochastic Microscopic Finsler Cosmology,''
  Eur.\ Phys.\ J.\ C {\bf 72}, 1956 (2012).  % doi:10.1140/epjc/s10052-012-1956-7  [arXiv:1012.4094 [hep-ph]].
  %%CITATION = doi:10.1140/epjc/s10052-012-1956-7;%%
 
\bibitem{Witten2002wb}  See, e.g., E.~Witten,
  ``Comments on string theory,''
  arXiv:hep-th/0212247 (2002), and references therein.  
  
%\cite{Abel:2015tca}
\bibitem{kehagias} S.~Kasuya and M.~Kawasaki,
  %``A New type of stable Q balls in the gauge mediated SUSY breaking,''
  Phys.\ Rev.\ Lett.\  {\bf 85}, 2677 (2000);
%  doi:10.1103/PhysRevLett.85.2677
%  [hep-ph/0006128].
  %%CITATION = doi:10.1103/PhysRevLett.85.2677;%%
S.~Kasuya, M.~Kawasaki and T.~T.~Yanagida,
  %``IceCube potential for detecting Q-ball dark matter in gauge mediation,''
  PTEP {\bf 2015}, no. 5, 053B02 (2015);
%  doi:10.1093/ptep/ptv056
%  [arXiv:1502.00715 [hep-ph]]
  %%CITATION = doi:10.1093/ptep/ptv056;%%
S.~Abel and A.~Kehagias,
  %``Q-branes,''
  JHEP {\bf 1511}, 096 (2015).
%  doi:10.1007/JHEP11(2015)096
%  [arXiv:1507.04557 [hep-th]].
  %%CITATION = doi:10.1007/JHEP11(2015)096;%% 

\end{thebibliography}
\end{document}